\documentclass[fleqn,useAMS,usenatbib]{mnras}

\usepackage{newtxtext,newtxmath}
\usepackage[T1]{fontenc}
\usepackage{ae,aecompl}

\usepackage{pdflscape}
\usepackage{graphicx}	

\newcommand{\kms}{\hbox{${\rm km\;s}^{-1}$}}

\newcommand{\hi}{H~\textsc{i}}   
\newcommand{\ha}{H$\alpha$}

\newcommand{\emax}{\ensuremath{\epsilon_{\mathrm{max}}}}

\newcommand{\amax}{\ensuremath{a_{\epsilon}}}

\newcommand{\lbar}{\ensuremath{L_{\mathrm{bar}}}}

\newcommand{\avis}{\ensuremath{a_{\mathrm{vis}}}}

\newcommand{\Msun}{\ensuremath{M_{\sun}}}
\newcommand{\Mstar}{\ensuremath{M_{\star}}}
\newcommand{\logmstar}{\ensuremath{\log \, (M_{\star}/M_{\sun})}}
\newcommand{\sfourg}{S\ensuremath{^{4}}G}
\newcommand{\sfourgplus}{S\ensuremath{^{4}}G+ETG}

\newcommand{\logmstarshort}{\ensuremath{\log \, M_{\star}}}
\newcommand{\logamax}{\ensuremath{\log \, a_{\epsilon}}}
\newcommand{\pbeta}{\ensuremath{P_{\beta=0}}}

\newcommand{\msepred}{\ensuremath{\mathrm{MSE}_{\mathrm{pred}}}}

\usepackage{xcolor}

\title[Frequencies and Sizes of Inner Structures in Bars]{The Frequency
and Sizes of Inner Bars and Nuclear Rings in Barred Galaxies and Their
Dependence on Galaxy Properties}

\author[Peter Erwin]{Peter Erwin$^{1,2}$\thanks{E-mail: 
erwin@mpe.mpg.de}\\
$^{1}$Max-Planck-Institute for Extraterrestrial Physics, Giessenbachstr., 
85748 Garching, Germany\\
$^{2}$Universit\"{a}ts-Sternwarte M\"{u}nchen, Scheinerstrasse 1, D-81679 M\"{u}nchen, Germany}

\date{Accepted XXX. Received YYY; in original form ZZZ}

\pubyear{2023}

\begin{document}
\label{firstpage}
\pagerange{\pageref{firstpage}--\pageref{lastpage}}
\maketitle

\begin{abstract} 

Using a volume- and mass-limited ($D < 30$ Mpc, $\logmstar \geq 9.75$)
sample of 155 barred S0--Sd galaxies, I determine the fraction with
secondary structures within their bars. Some $20 \pm 3$\% have a
separate inner bar, making them double-barred; an identical fraction
have nuclear rings, with $11^{+3}_{-2}$\% hosting both. The inner-bar
frequency is a strong, monotonic function of stellar mass: only
$4^{+3}_{-2}$\% of barred galaxies with $\logmstar = 9.75$--10.25 are
double-barred, while $47 \pm 8$\% of those with $\logmstar > 10.5$ are.
The nuclear-ring frequency is a strong function of absolute bar size:
only $1^{+2}_{-1}$\% of bars with semi-major axes $< 2$~kpc have nuclear
rings, while $39^{+6}_{-5}$\% of larger bars do. Both inner bars and
nuclear rings are absent in very late-type (Scd--Sd) galaxies.

Inner bar size correlates with galaxy stellar mass, but is clearly
offset to smaller sizes from the main population of bars. This makes it
possible to define ``nuclear bars'' in a consistent fashion, based on
stellar mass. There are eight single-barred galaxies where the bars are
nuclear-bar-sized; some of these may be systems where an outer bar
failed to form, or previously double-barred galaxies where the outer bar
has dissolved. Inner bar size is even more tightly correlated with host
bar size, which is likely the primary driver. In contrast, nuclear ring
size is only weakly correlated with galaxy mass or bar size, with more
scatter in size than is true of inner bars. 

\end{abstract}


\section{Introduction} 

The interiors of galactic bars can sometimes be rather complex and
dramatic places. Instances of double-barred galaxies -- where a second,
smaller bar is nested inside a large-scale bar, usually with a different
orientation -- have been known since the 1970s
\citep[e.g.,][]{dev75-db,sandage79}. Examples of nuclear rings -- where
a ring or pseudoring, often actively star-forming or even starbursting,
encircles the nucleus inside the bar -- have been known even longer
\citep[see references in][]{buta96}

Double bars first attracted general interest for their possible role
in fuelling AGNs. \citet{shlosman89} argued that while large-scale bars
were known to drive gas to smaller radii, this inflow might stall near
the centers of such bars, at radii of a few hundred pc. The presence of
a separate inner bar -- perhaps formed out of the disc of gas funneled
inward by the large-scale bar -- could then drive gas inward to smaller
scales, where other instabilities might take over. Observational tests
of this idea \citep[e.g.,][]{erwin02,laine02} have generally been
inconclusive, however. 

More recently, the simulations of \citet{du17} suggested that inner bars
were more vulnerable to destruction by central supermassive black holes
than were large-scale bars, and consequently disrupted inner bars
might be the origin of central (classical) bulges in some galaxies
\citep[see also][]{guo20,nakatsuno23}.

There have even been occasional suggestions that the Milky Way itself might
be a double-barred galaxy
\citep[e.g.,][]{alard01,nishiyama05,nishiyama06,namekata09}, though the
edge-on nature of the Galaxy disk makes it difficult to determine and to
distinguish between potentially similar-sized, flattened
\textit{axisymmetric} structures such as a nuclear stellar disk
\citep[e.g.,][]{gerhard12,valenti16}.

Nuclear rings have probably been the focus of more research, in part
because the presence of significant star formation so close to the
center of a galaxy suggests a role in building up the central structure
of galaxies in the form of nuclear discs or pseudobulges. The most
obvious -- and earliest known -- examples are naturally those rings with
active star formation, though smooth rings of old stars -- possibly the
faded, fossil remnants of star-forming rings -- also exist
\citep[e.g.,][]{erwin01-nr}, as do dusty, gaseous rings without ongoing
star formation.  As with inner bars, it is possible that the Milky Way
hosts a nuclear ring within its bar as well
\citep[e.g.,][]{molinari11,henshaw16,sormani18a}.

Currently we have very little idea how common double bars really are;
the only surveys to date have either focused on a narrow range of Hubble
types \citep[e.g.,][]{erwin02} or have looked at potentially biased
subsets such as Seyfert galaxies \citep[e.g.,][]{laine02}. The situation
for nuclear rings is somewhat better, with surveys such as
\citet{knapen05} and the catalog of \citet{comeron10}. (Some nuclear
rings and inner bars are listed in the comprehensive \sfourg{} analysis
of \citealt{herrera-endoqui15}, but these are limited by the low
resolution of the IRAC1 images.) Recent studies of bars have shown that
some substructures of bars -- in particular, the boxy/peanut-shaped (BP)
bulges that are the vertically thickened inner parts of some bars --
have very strong dependences on galaxy stellar mass
\citep[e.g.,][]{erwin17,li17,kruk19,erwin23}. So we might wonder if inner
bars or nuclear rings might have a similar dependence on stellar mass.
This paper is meant to investigate this and related questions by
determining the frequency (and sizes) of inner bars and nuclear rings in
an unbiased, distance- and mass-limited sample of barred galaxies.

\section{The Sample}\label{sec:sample} 

The foundation of the sample studied here is a survey of S0--Sa
barred galaxies presented in \citet{erwin02} and
\citet{erwin-sparke03}. This was later extended to include barred Sab and Sb
galaxies in \citet{erwin05} and \citet{erwin08}, with additional barred
galaxies identified among the corresponding (nominally) unbarred S0--Sb
galaxies by \citet{gutierrez11}. The original sample was defined to
include all northern ($\delta > -10\degr$) barred galaxies from the
Uppsala General Catalog \citep{ugc} with heliocentric redshifts $\leq
2000$ \kms{} (roughly speaking, $D < 28$ Mpc), RC3 major-axis diameters
$D_{25} \geq 2.0$\arcmin, and RC3 axis ratios $a/b \leq 2.0$.

In this paper, I extend the sample concept to cover the full range of
bars in regular disks (Hubble types S0--Sd). I also redefine it to have
rigorous distance and stellar-mass limits, and drop the original
angular-diameter limit. The starting sample is now that defined in the
\textit{Spitzer} Survey of Stellar Structure in Galaxies
\citep[\sfourg;][]{sheth10} along with the corresponding Early-Type
Galaxies Extension \citep{watkins22}, which corrects for the absence of
most elliptical and S0 galaxies in the original \sfourg. (I refer to the
combined sample of \citealt{sheth10} and \citealt{watkins22} as
\sfourgplus.) More specifically, I start with all S0--Sd galaxies in
\sfourgplus{} (using their optical Hubble types, which come from
HyperLEDA),\footnote{I exclude two galaxies (NGC~4116 and NGC~4496A)
with RC3 types of Sdm and Sm.} with $D < 30$ Mpc, stellar masses
$\logmstar > 9.75$,\footnote{The lower limit is meant to make the
most of previous analyses of local barred galaxies
\citep[e.g.,][]{erwin-sparke03,erwin11}, which extended to masses
slightly below $\logmstar \sim 10$, as well as to keep the fraction of
galaxies observed with \textit{HST} reasonably high. ($\sim 50$\% of
sample galaxies with $\logmstar = 9.75$--10 have \textit{HST} data,
while only 28\% of galaxies that would meet the sample criteria, but
with $\logmstar = 9.0$--9.75, do.)} and RC3 axis ratios $a/b \leq 2.0$
(corresponding to inclinations $i \la 62\degr$). For consistency with
previous analyses, and to take advantage of the greater availability of
imaging for northern galaxies, I retain the $\delta > -10\degr$ limit.

This yielded an initial sample of 279 galaxies. A total of 27 galaxies
were discarded for a variety of reasons (e.g., being too highly
inclined, despite their RC3 axis ratios or being strongly distorted by
interactions or mergers); these rejections are summarized in
Appendix~\ref{app:discarded}.

I also updated the galaxy distance estimates. Distances in the
\sfourgplus{} databases are mostly taken from NED circa 2015
\citep[e.g.,][]{munoz-mateos15}. A handful of galaxies in the initial
sample have more recent direct measurements (Cepheids, TRGB, SBF); in
addition, I used group distance measurements from
\citet{kourkchi-tully17} for galaxies in groups where at least three
group members had independent distance measurements. These newer
distances were used only if they differed from the \sfourgplus{} values
by more than 10\%. This led to ten additional galaxies being discarded
for having updated distances $> 30$ Mpc, and one for having a revised
stellar-mass estimate below the sample cutoff (see
Appendix~\ref{app:discarded}). The parent sample then has a total of 242
galaxies; in the next section I describe how this was reduced to the
final sample of barred galaxies.


\begin{table*}
\hspace*{-25mm}
\begin{minipage}{150mm}
    \caption{Barred Galaxy Sample}
    \label{tab:sample}
    \begin{tabular}{llrrrrrrlrrrrrrr}
\hline
Name & Type & T     & D        & $\log M_{\star}$ & Disc PA & $i$   & Source & rot & Bar PA &  \amax     & \lbar     & \emax & DB?  & NR?  & FWHM \\
     &      &       & (Mpc)    & ($\Msun$)        & (deg)   & (deg) &        &     & (deg)  &  (\arcsec) & (\arcsec) &        &      &      & (\arcsec) \\
(1)  & (2)  & (3)   & (4)      & (5)              & (6)     & (7)   & (8)    & (9) & (10)   &  (11)       & (12)       & (13)   & (14) & (15) & (16) \\
\hline
IC 676   & (R)SB(r)$0^{+}$          & $-1.2$ &   20.0            &  9.832 &   15 &   47 & 4  & $-$? &  164 & 13   & 18   & 0.72 &   &   & 0.15 \\
IC 1067  & SB(s)b                   & $ 3.0$ &   21.9            &  9.913 &  120 &   44 & 4  & $-$  &  151 & 19   & 19   & 0.64 &   &   & 0.74 \\
NGC 514  & SAB(rs)c                 & $ 5.2$ &   29.4            & 10.390 &  104 &   48 & 4  & +    &  154 & 10   & 12   & 0.32 &   &   & 0.15 \\
NGC 600  & (R')SB(rs)d              & $ 7.0$ &   22.9            &  9.793 &   20 &   36 & 3  & +    &   18 & 15   & 16   & 0.68 &   &   & 0.07 \\
NGC 718  & SAB(s)a                  & $ 1.0$ &   21.4            & 10.283 &    5 &   30 & 4  & +    &  152 & 20   & 30   & 0.44 & Y & Y & 1.00 \\
NGC 864  & SAB(rs)c                 & $ 5.1$ &   18.8            & 10.184 &   23 &   43 & 4  & +    &   96 & 18   & 23   & 0.50 &   & Y & 0.10 \\

\hline
\end{tabular}

\medskip

Barred galaxy sample. (1) Galaxy name. (2) Hubble type (Leda). (3) Numerical
Hubble type. (4) Distance (from \sfourgplus{}, except as noted). (5) Stellar mass (from
\sfourgplus, updated for changed distances). (6) Position angle of disc
line of nodes. (7) Inclination of disc. (8) Source for disc orientation
(1 = SDSS images; 2 = SGA images; 3 = IRAC1 image; 4 = see notes in
Appendix~\ref{app:barred}). (9) Sense of galaxy rotation (+ = counter-clockwise,
$-$ = clockwise) from spiral arms, if available.
(10) Position angle of bar (outer bar in
double-barred galaxies). (11) Bar size \amax. (12) Bar size \lbar. (13)
Bar maximum ellipticity \emax. (14) Galaxy is double-barred. (15) Galaxy has nuclear ring.
(16) Full-width-half-maximum of PSF of best available image used to
determine DB/NR presence/absence. Notes on alternate distance sources:
(a) Cepheids from \citet{saha06}; (b) SBF from \citet{tonry01}, with correction from \citet{mei05}; 
(c) Group distance from \citet{kourkchi-tully17}; (d) Cepheids from \citet{yuan20}; 
(e) Virgo Cluster distance from \citet{mei07}; (f) SBF from \citet{cantiello18}; 
(g) Cepheids from \citet{riess16}. The full table is available in the online version of
this paper; I show a representative sample here.

\end{minipage}
\end{table*}

\subsection{Constructing the Final Sample: Identification of Barred Galaxies}\label{sec:id-barred} 

The vast majority of galaxies in the final (barred) sample have been
classified as barred in \sfourg{} or its extension. However, it is worth
carefully analyzing all the individual galaxies in order to exclude
unbarred galaxies erroneously classified as barred, and to include
barred galaxies erroneously classified as unbarred. This was done by
visual inspection of the \textit{Spitzer} IRAC1 (3.6\micron) images that
exist for all the galaxies, as well as any available archival
\textit{Hubble Space Telescope} (\textit{HST}) images and ground-based
images from the Sloan Digital Sky Survey (SDSS; mostly Data Release 7
\citealt{abazajian09}) and the Siena Galaxy Atlas
\citep[SGA;][]{moustakas23}.

The application of this analysis to many of the S0--Sb galaxies in the
parent sample has been previously discussed in \citet{erwin-sparke03},
\citet{erwin05} and \citet{gutierrez11}; the latter noted the existence
of bars in twelve nominally ``unbarred'' galaxies (their Table~7).

The following galaxies classified as unbarred in \sfourgplus{} (either
in \citealt{buta15} and \citealt{herrera-endoqui15} for \sfourg{}
galaxies or via the ``bar family'' classification in
\citealt{watkins22}) proved to have bars: NGC~514, NGC~1068 (see
\citealt{erwin04}), NGC~2776, NGC~2964, NGC~3031 \citep{gutierrez11},
NGC~3599 \citep{gutierrez11}, NGC~3658, NGC~3982, NGC~3998
\citep{gutierrez11}, NGC~4041, NGC~4203 \citep{erwin-sparke03},
NGC~4377, NGC~4736 (see \citealt{erwin04}), NGC~4750, NGC 4941, NGC
5194/M51a \citep[e.g.,][]{menendez-delmestre07}, and NGC~5300.
Conversely, I found no convincing evidence for bars in NGC~3430,
NGC~3949, NGC~4351, NGC~5665, NGC~5958, or NGC~5963.

The final result is a sample of 155 barred galaxies, with 40 S0s and 115
spirals. (The unbarred galaxies amount to 86: 16 S0s and 70 spirals.)
These galaxies, along with basic parameters and measurements of the
(outer or sole) bars, are listed in Table~\ref{tab:sample}.

\begin{figure*}
\begin{center}
\hspace*{-5mm}
\includegraphics[scale=0.85]{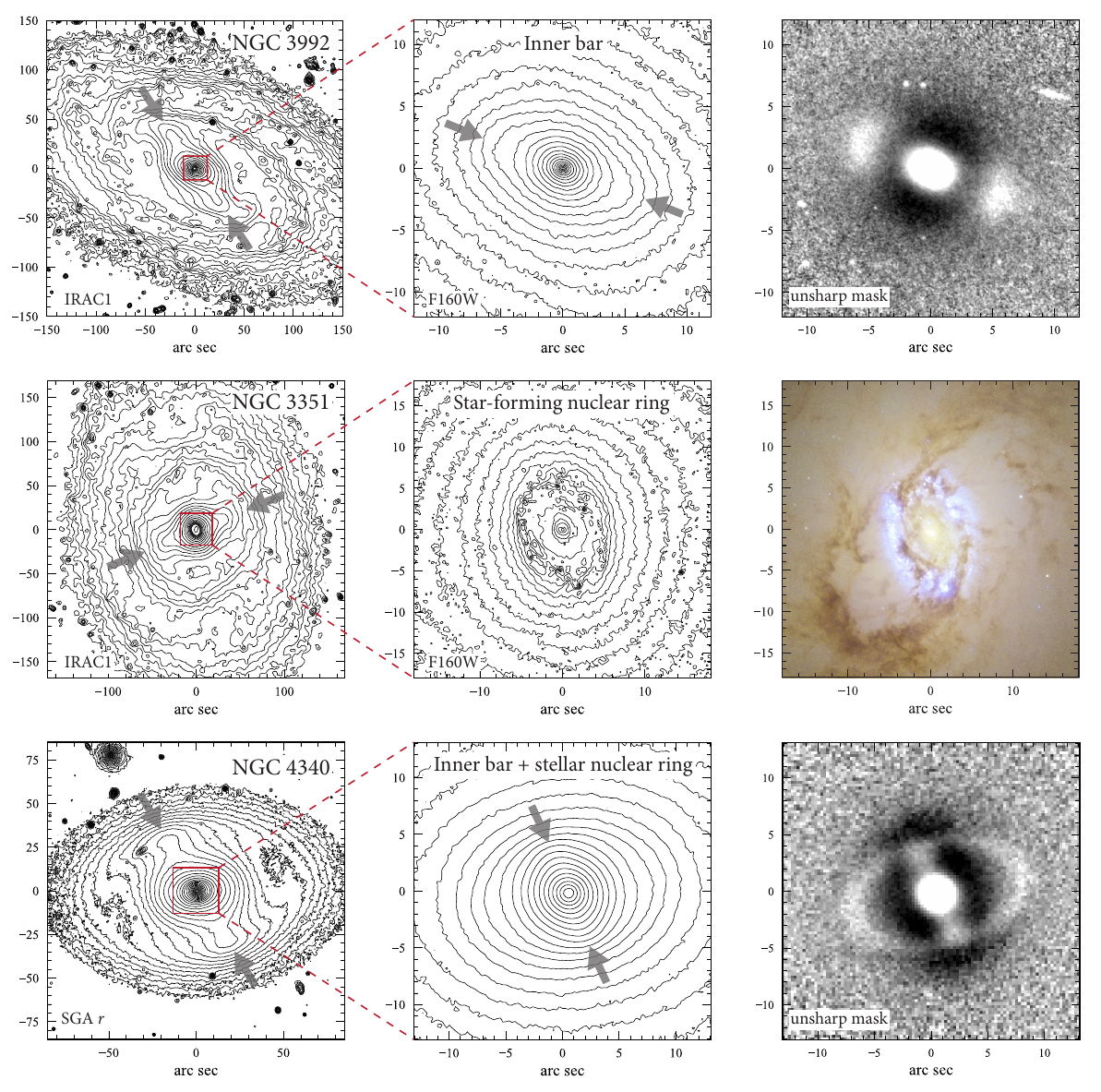}
\end{center}

\caption{Examples of barred galaxies with inner bars and nuclear rings.
Contour plots are logarithmically spaced surface brightness; N is up and
E is to the left in all panels; large grey arrows indicate bars.
\textbf{Top row:} SBbc galaxy NGC~3992 (M109), with an inner bar. Left
panel = \sfourg{} IRAC1 image; middle panel = \textit{HST} WFC3-IR F160W image
(Proposal ID 15323, PI Walsh); right panel = unsharp mask of F160W
image. \textbf{Middle row:} SBb galaxy NGC~3351 (M95), with a
star-forming nuclear ring; left panel = \sfourg{} IRAC1 image; middle
panel = \textit{HST} WFC3-IR F160W image (Proposal ID 15133, PI Erwin);
right panel = WFC3-UVIS UV and optical composite image (F275W, F336W,
F438W, F555W, F814W; cropped from original image of ESA/Hubble \& NASA).
\textbf{Bottom row:} SB0 galaxy NGC~4340, with both an inner bar and a
(stellar, non-star-forming) nuclear ring. Left and middle panels: SGA
$r$ image; right panel: unsharp mask of same.}
\label{fig:examples}

\end{figure*}

\subsection{Measurements of Disks and Bars} 

In order to study the sizes of bars and rings, it is usually necessary
to correct for projection effects, which will shorten any structures
whose major axes are not parallel to the major axis of the disc. This
requires knowing the orientation of the disc: its inclination and the
position angle of its major axis. These are listed for all galaxies in
Table~\ref{tab:sample}. In the vast majority of cases, the disc
orientations come from fitting ellipses to the isophotes of the galaxy
well outside the bar, using the \textsc{iraf} task \texttt{ellipse},
which implements the algorithm of \citet{jedrzejewski87}, and assuming
that the outer disc (i.e., well outside any bar) is intrinsically circular with a
vertical-to-radial axis ratio of $c/a = 0.2$. The sources (generally
either SDSS, SGA, or IRAC1 images) are listed for individual galaxies 
in the table. Exceptions -- generally cases where, e.g., literature
measurements based on gas velocity fields provided more reliable
estimates than outer isophotes with low S/N or distortions due to strong
spiral arms -- are discussed in Appendix~\ref{app:barred}.

Ellipse fitting is also used to measure the sizes, shapes, and
orientations of bars, along with visual inspection to ensure that peaks
in the ellipticity profile correspond to a bar rather than to dust
extinction, spiral arms, or rings. \citet{erwin04} and especially
\citet{erwin05} provide detailed discussions and examples of this. For consistency
with previously published measurements -- including the bar measurements
for 62 galaxies (17 of them with double bars) \citep{erwin-sparke03,
erwin04,erwin08,gutierrez11} in this sample -- I use \amax{} and
\lbar{}. The smaller measure, \amax{}, is the semi-major axis of maximum
ellipticity in the bar region; sometimes this can be an extremum in
the position angle instead, as can be the case when a bar is oriented
near the minor axis of a highly inclined galaxy and produces a
\textit{minimum} in the ellipticity (see, for example, the cases of
NGC~2787, NGC~2880, NGC~3412, NGC~3489, and NGC~4143 in
\citealt{erwin-sparke03}). Following \citet{erwin05}, the larger size,
\lbar{}, is intended to be an upper-limit measurement, and is usually
the minimum of three measurements: the semi-major axis of the first
ellipticity \textit{minimum} outside the bar; the semi-major axis where
the isophote PA differs by $> 10\degr$ from the bar PA; and the
semi-major axis of any spiral arms or rings that visibly cross the end
of the bar. Finally, the PA of the bar is taken from a combination of
the ellipse fits, visual inspection, and unsharp masking \citep[see the
discussion in][]{erwin-sparke03}. The resulting measurements are
included in Table~\ref{tab:sample}.

\section{Identification of Inner Bars and Nuclear Rings} 

The process of identifying inner bars and nuclear rings relies on a
combination of visual inspection, fitting ellipses to isophotes, and
unsharp masking, applied to optical and near-IR images. Image sources
included \textit{Spitzer} IRAC1 images from \sfourgplus; optical images
from SGA and SDSS; and archival images from \textit{HST}.

\textbf{A note on terminology and definitions:} For galaxies with
two bars (``double-barred'' galaxies), the large bar is the ``outer
bar'' (also known as the ``primary bar'' in the literature) and the
small bar is the ``inner bar'' (a.k.a. ``secondary bar''). (The question
of whether some bars can be called ``nuclear bars'' will be taken up in
Section~\ref{sec:nuclear-bars}.) In galaxies with only one bar, the bar
is ``single'' or ``sole''. ``Nuclear rings'' are defined to be rings (or
spiral pseudorings) located \textit{inside} a bar (or inside the outer
bar in double-barred systems).

Inner bars are found and measured using the methodology outlined in the
preceding section. Unsharp masking is especially important for
confirming the presence of weak inner bars and for avoiding confusion
due to nuclear spirals, nuclear disks, and nuclear rings, which can on
occasion produce similar ellipse-fit signatures \citep[see the examples
in][]{erwin04}. Figure~\ref{fig:examples} shows how inner bars can be
identified via visual inspection and unsharp masking (NGC~3992 and
NGC~4340). The example of NGC~4340 (bottom panels of the figure)
highlights the utility of unsharp masks, which in this case show both an
inner bar \textit{and} a stellar nuclear ring surrounding the inner bar.
 
The identification of nuclear rings was made using a combination of
visual inspection of images (especially optical images, where both star
formation and dust extinction are more prominent than they are in
near-IR images -- see the example of NGC~3351 in the middle panels of
Figure~\ref{fig:examples}) and unsharp masks; the latter are especially
useful for identifying stellar nuclear rings (see below).

I sub-classify the nuclear rings a similar fashion to
\citet{erwin-sparke03}, into star-forming, purely stellar, and
purely dusty. The first category includes both actively star-forming
rings and rings that are distinctly bluer than the surrounding bar
(suggestive of relatively recent star formation). These are usually easy
to identify due to their blue colours, patchy nature (the combination of
recent sites of localized star formation and dust extinction), and \ha{}
emission; see, e.g., \citet{comeron10} for numerous examples.\footnote{Note,
however, that Comer{\'o}n et al.'s ``star-forming'' nuclear ring class includes
the passive, post-star-forming features I call stellar nuclear rings.}

The stellar nuclear rings are smooth circular or oval structures, with
colours similar to the surrounding bar, suggesting that they are the
fossil remnants of past star-forming rings. These are often best
revealed in unsharp masks (Figure~\ref{fig:examples}; see also
\citealt{erwin99,erwin01-nr}). 

Dusty nuclear rings are cases where dust lanes form ringlike or
pseudoring structures, \textit{without} evidence for accompanying star
formation (or a passive stellar nuclear ring, though these might be
obscured by the dust lanes). The assumption is that the rings are more
properly \textit{gaseous} rings, made visible in the optical and near-IR
by the accompanying dust. These are relatively rare; I find only three
clear examples (NGC~2859, NGC~3489, and NGC~5377).


\begin{table}
\caption{Inner-bar Measurements for Double-Barred Galaxes}
\label{tab:db}
\begin{tabular}{@{}lrrrr}
\hline
Name & Bar PA &  \amax     & \lbar     & \emax \\
     & (deg)  &  (\arcsec) & (\arcsec) & \\
(1)  & (2)    & (3)         & (4)        & (5) \\
\hline
NGC 718  & 15 & 1.6 & 2.2 & 0.19 \\
NGC 1068 & 47 & 15 & 17 & 0.45 \\
NGC 2681 & 20 & 1.7 & 3.3 & 0.26 \\
NGC 2859 & 62 & 4.1 & 6.2 & 0.31 \\
NGC 2950 & 85 & 3.2 & 3.9 & 0.33 \\
NGC 3031 & 140 & 17 & 27 & 0.39 \\

\hline
\end{tabular}

\medskip

Measurements of inner bars for double-barred galaxies. (1) Galaxy name.
(2) Position angle of inner bar. (3) Bar size \amax. (4) Bar size \lbar.
(5) Bar maximum ellipticity \emax. The full table is available in the
online version of this paper; I show a representative sample here.
\end{table}


\begin{table}
\caption{Nuclear Ring Measurements}
\label{tab:nr}
\begin{tabular}{@{}lrrrl}
\hline
Name & NR PA &  NR $a$     & $\epsilon$ & Class \\
     & (deg) &  (\arcsec)  &            & \\
(1)  & (2)   & (3)         & (4)        & (5) \\
\hline
NGC 718  & 50 & 3.3 & 0.15 & blue \\
NGC 864  & 110 & 0.8 & 0.13 & star-forming \\
NGC 936  & 130 & 8.5 & 0.23 & stellar \\
NGC 1068 & 50 & 16 & 0.32 & star-forming \\
NGC 2608 & 45 & 1.1 & 0.27 & star-forming \\
NGC 2859 & 60 & 7.0 & 0.20 & dust \\

\hline
\end{tabular}

\medskip

Measurements of nuclear rings. (1) Galaxy name. (2) Position angle of nuclear ring. 
(3) Semi-major axis. (4) Ellipticity. (5) Nuclear-ring class. The full 
table is available in the online version of this paper; I show a representative sample here.
\end{table}

\section{Frequencies of Inner Bars and Nuclear Rings}\label{sec:frequencies} 

There are a total of 31 inner bars in the sample of 155 barred galaxies,
so $20 \pm 3$\% of the barred galaxies -- or $13 \pm 2$\% of all the
galaxies, barred and unbarred -- are double-barred. Their parameters are
listed in Table~\ref{tab:db}. There are also a total of 31 nuclear rings
in the sample, with parameters listed in Table~\ref{tab:nr}; the
nuclear-ring fraction is thus identical to the double-bar fraction.
(Which galaxies are double-barred or nuclear-ring hosts is also
indicated in Table~\ref{tab:sample}.) Seventeen of the galaxies host
\textit{both} an inner bar \textit{and} a nuclear ring ($11^{+3}_{-2}$\%
of the barred galaxies, or $7^{+2}_{-1}$\% of all the disc galaxies).

A potentially more interesting question than the raw numbers or frequencies in the
sample is: how do the fractions of bars with inner bars or nuclear rings
depend on galaxy properties? In the following sections, I look at how
this might depend on galaxy stellar mass -- since the presence of both
large-scale bars and B/P bulges inside bars depends strongly on stellar
mass \citep[e.g.,][]{erwin17,li17,erwin18,erwin23} -- and also on the
size of outer bars (or the only bar in the case of single-barred
galaxies) and on Hubble type. I first focus on possible trends, and then
use logistic regression to quantify the trends and identify which galaxy
parameters might be most important for determining whether a galaxy is
double-barred or has a nuclear ring. The stellar masses are all taken
from \sfourgplus{} \citep[][]{munoz-mateos15,watkins22}, except for
those galaxies where an updated distance is used
(Section~\ref{sec:sample}), for which I recomputed the masses
appropriately.

\begin{figure}
\includegraphics[scale=0.68]{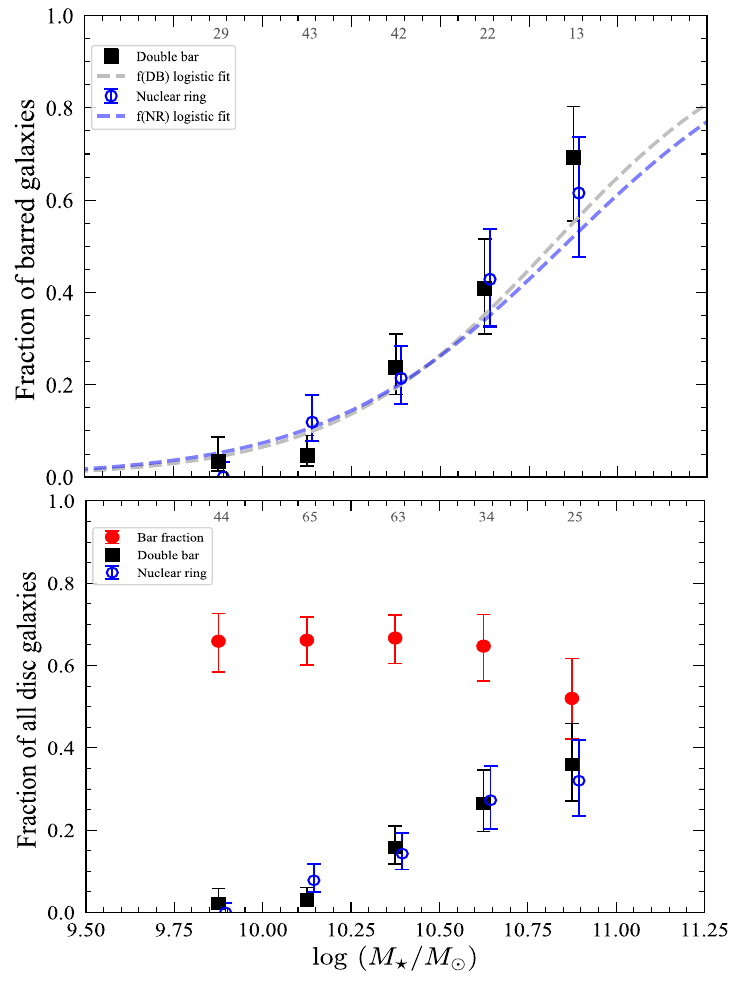}

\caption{Frequency of inner bars (black squares) and nuclear rings (blue
circles) in barred S0--Sd galaxies (top) and all S0--Sd galaxies
(bottom) as a function of galaxy stellar mass. Dashed curves in the top
panel show logistic fits (probability of a galaxy to have the indicated
structure) to the underlying data.
Red circles in the bottom panel show the fraction of galaxies with at
least one bar. The small grey numbers along the tops of each panel give
the number of galaxies in each bin. There is a
clear trend of both inner bars and nuclear rings becoming more common
with increasing stellar mass.}

\label{fig:fDB-NR-vs-logMstar}
\end{figure}

\subsection{Inner Bars} 

Figure~\ref{fig:fDB-NR-vs-logMstar} shows the frequency of inner bars as
a function of galaxy stellar mass for barred galaxies (top panel) and
for all galaxies, barred and unbarred (bottom panel).\footnote{Note that
overall bar-fraction trend in the bottom panel (red circles) declines
less steeply with stellar mass than was true for the spiral-only sample
analyzed in \citet{erwin18}. This appears to be due to the inclusion of
S0s in this paper's sample; see Erwin (in prep).} What is quite striking
is the very clear and strong mass dependence: inner bars are
essentially \textit{absent} for stellar masses $\logmstar \la 10$
and then increase steeply in frequency to higher masses, reaching a
fraction of $47 \pm 8$\% for $\logmstar > 10.5$ for barred galaxies, or
$26^{+6}_{-5}$\% when considering all S0--Sd galaxies. The lowest-mass
double-barred galaxy is NGC~5770, with $\logmstar = 9.94$.

\begin{figure}
\hspace*{-4.5mm}\includegraphics[scale=0.47]{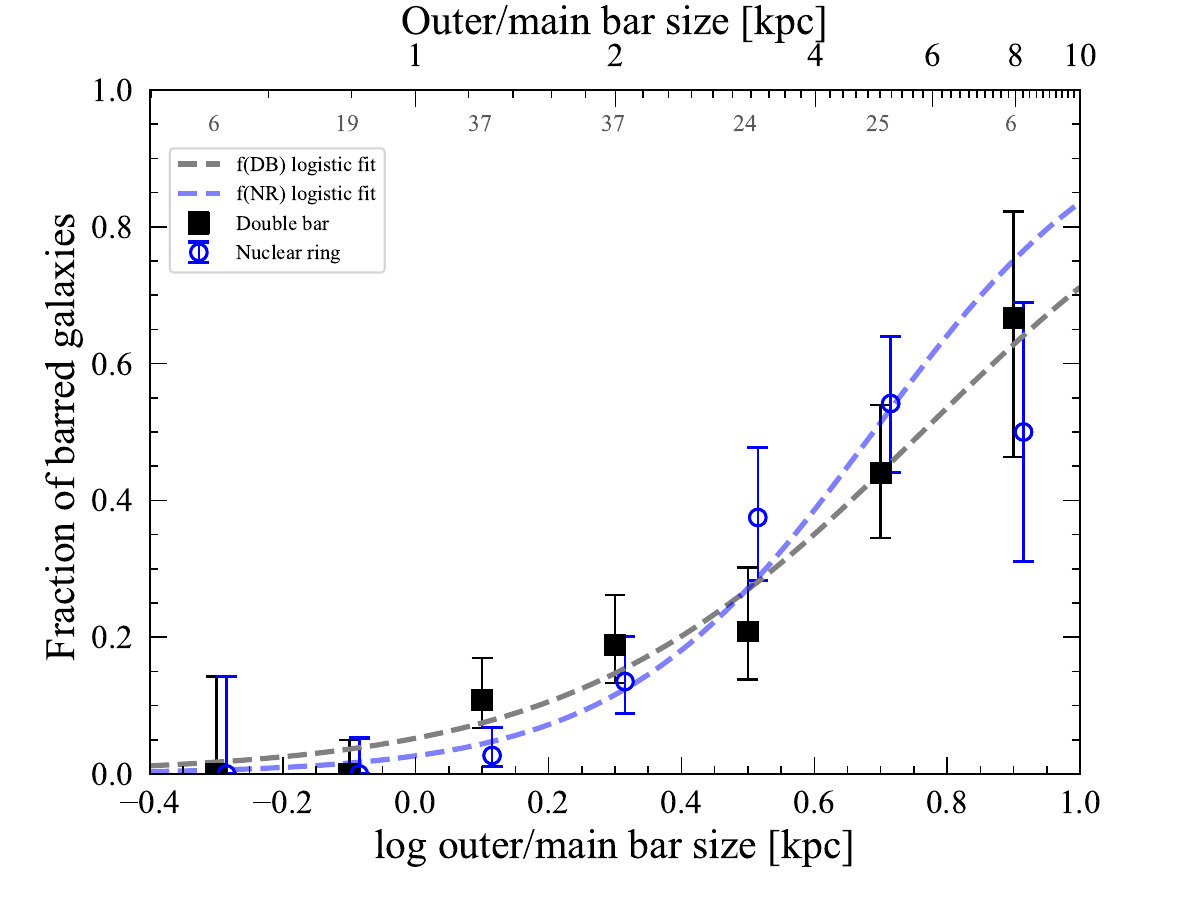}

\caption{Fraction of double bars (black squares) and nuclear rings (open
blue circles) for barred galaxies as a function of outer/single bar size
(\amax) in kpc. The dashed grey and blue lines are logistic fits for the
double-bar (DB) and nuclear-ring (NR) fractions, respectively. Numbers
along the top of the plot show the number of galaxies in each
(logarithmically spaced) bin.} \label{fig:fDB-NR-vs-logbarsize}

\end{figure}

In Figure~\ref{fig:fDB-NR-vs-logbarsize}, we can see that inner-bar
frequency is \textit{also} a strong function of (outer) bar size: the
larger a (large-scale) bar in a galaxy, the higher the chance that it is
the outer bar of a double-bar system. No bars with sizes $< 1$ kpc have
inner bars, while almost half ($48 \pm 9$\%) of bars with $\amax > 4$
kpc have inner bars.

\begin{figure}
\hspace*{-4.5mm}\includegraphics[scale=0.47]{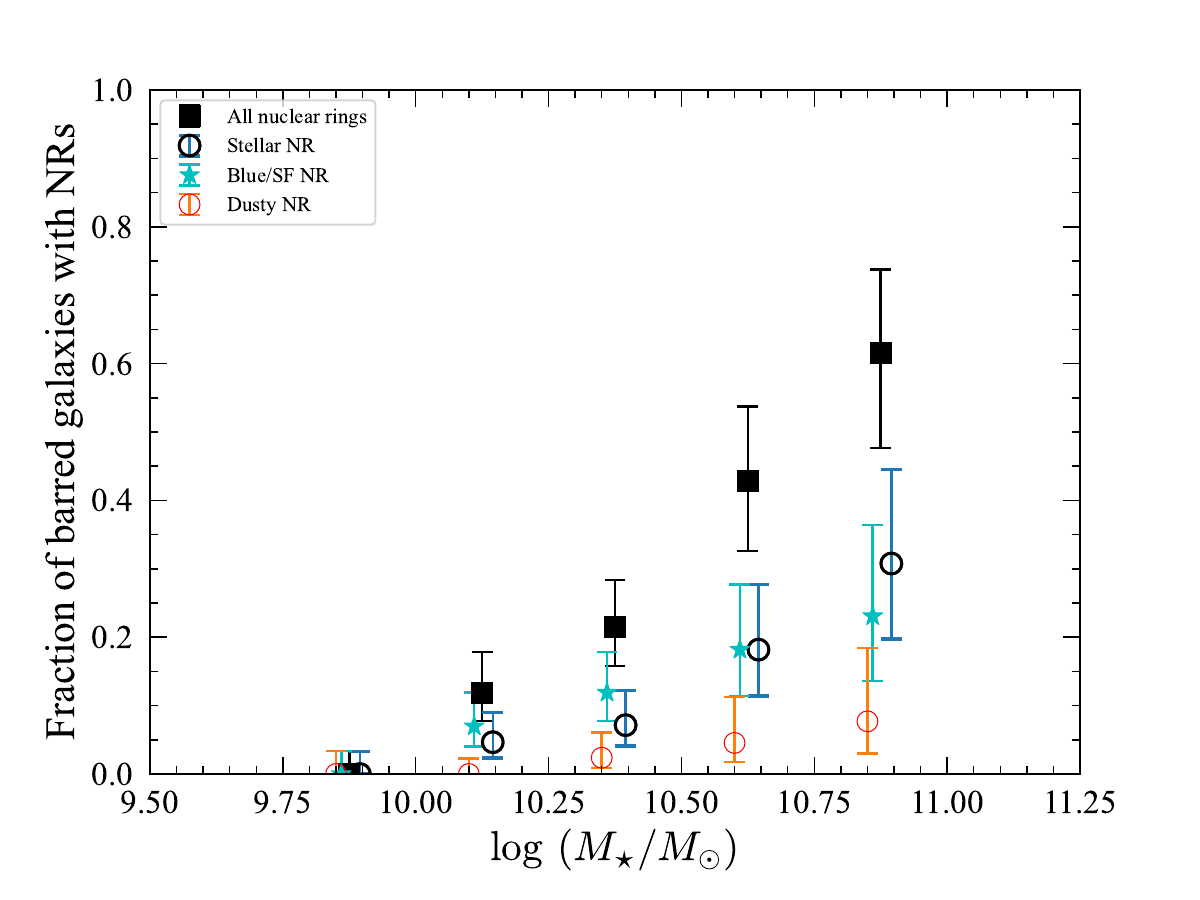}

\caption{Nuclear-ring fraction in barred galaxies for all nuclear rings
(black squares) and for different classes of nuclear ring: star-forming
(cyan stars), stellar (open black circles), and dust (open orange
circles).}
\label{fig:fNR-classes-vs-logmstar}

\end{figure}

\subsection{Nuclear Rings} 

Of the 31 nuclear rings in the sample, sixteen are actively forming
stars (or, in the case of NGC~718, are blue enough to suggest recent
star formation, so I include it in the ``star-forming'' category),
while thirteen are smooth, passive stellar rings. Only three are dusty
nuclear rings without evidence for star formation.

Figure~\ref{fig:fDB-NR-vs-logMstar} shows that the mass trend for
nuclear rings as a function of stellar mass is almost identical to that
for inner bars: nuclear rings are absent at low masses, and their
frequency increases monotonically to higher masses. The nuclear-ring
frequency at the high- and low-mass ends is almost identical to
that for inner bars (0\% for $\logmstar < 10$, $45 \pm 8$\% for
$\logmstar > 10.5$), with nuclear rings being marginally more common at
intermediate masses.

The breakdown for the three categories of nuclear ring is shown in
Figure~\ref{fig:fNR-classes-vs-logmstar}. Dusty nuclear rings are so rare
that no obvious trend is visible. Stellar and star-forming rings show
similar trends, increasing in frequency towards higher masses. There is
a suggestion that star-forming rings are more abundant in lower-mass
galaxies ($\logmstar \sim 10.0$--10.5), while stellar rings are more
abundant in the higher-mass galaxies ($\logmstar > 10.5$). Given the
uncertainties, however, it is not clear that this difference is
really meaningful.

Figure~\ref{fig:fDB-NR-vs-logbarsize} shows that nuclear-ring presence,
like inner-bar presence, is a strong function of bar size: only
$1.3^{+2.0}_{-0.8}$\% of bars with sizes $< 2$ kpc have nuclear rings,
while $39^{+6}_{-5}$\% of larger bars do. There is a hint that the
nuclear-ring frequency may depend on bar size more strongly than does
the inner-bar frequency; inner bars are more common than nuclear rings
at smaller bar sizes ($\amax \sim 1$--3 kpc).

\subsection{Double Bars \textit{and} Nuclear Rings} 

\begin{figure}
\hspace*{-4.5mm}\includegraphics[scale=0.47]{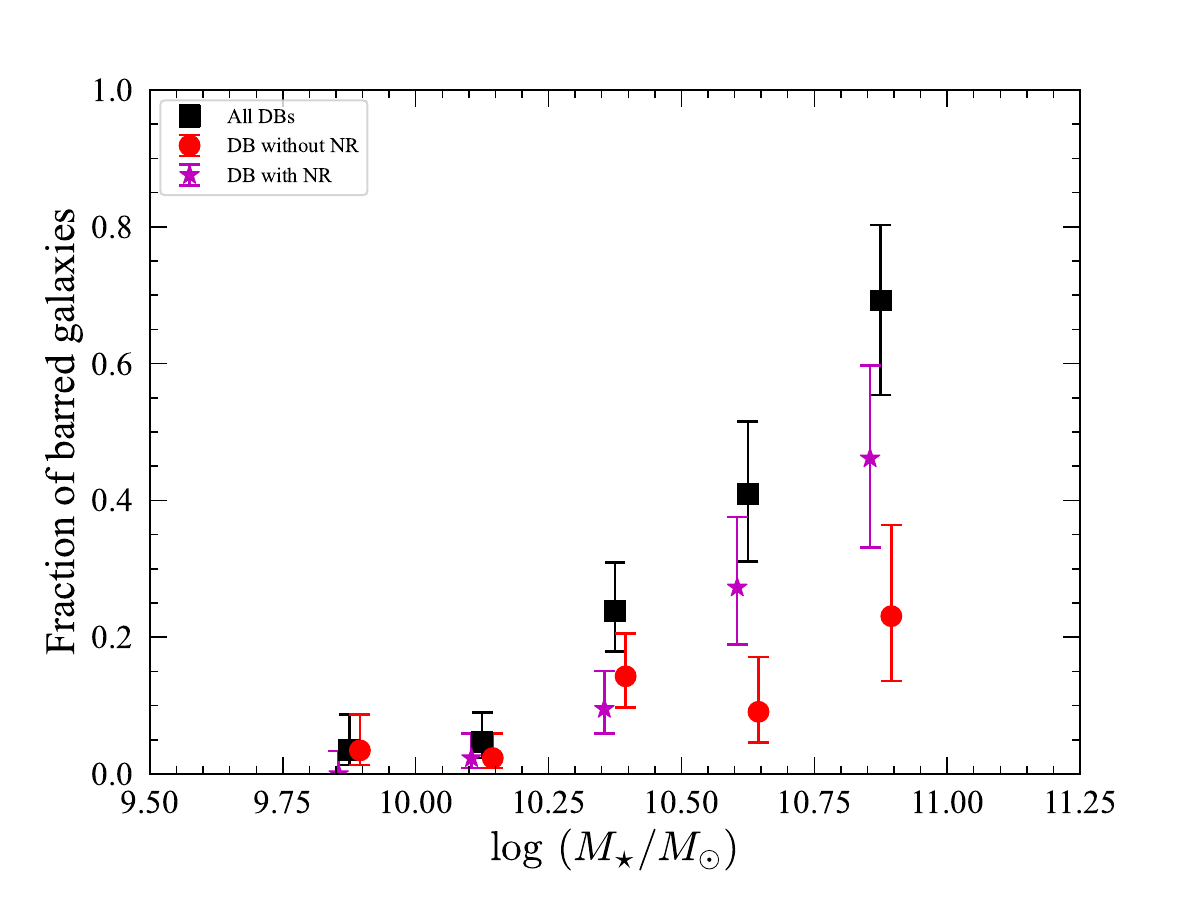}

\caption{Double-bar fraction for all barred galaxies (black squares) and
for those barred galaxies also hosting nuclear rings (magenta stars) or
those \textit{without} nuclear rings (red circles), as functions of
galaxy stellar mass.}
\label{fig:fDB-with-without-nr}

\end{figure}

The nearly identical frequencies and mass trends for inner bars and
nuclear rings suggests a strong association between the two structures
-- even that they might almost always appear together. Is this the case?
Figure~\ref{fig:fDB-with-without-nr} investigates this by comparing the
stellar-mass trends for all double bars (black squares) as well as for
double bars with (magenta stars) and without (red circles) accompanying
nuclear rings. In all cases we see a monotonic increase in frequency
with stellar mass. The main apparent trend is that in high-mass
galaxies, double bars are more likely to coexist with nuclear rings:
$65^{+10}_{-11}$\% of double bars in galaxies with $\logmstar >
10.5$ are accompanied by nuclear rings, versus only
$38^{+14}_{-12}$\% in lower-mass galaxies, although the
significance of this difference is marginal.

\subsection{Logistic Regression, Single- and Multi-variable}\label{sec:logistic} 

Since the trends of double-bar and nuclear-ring fraction show such
strong and monotonic trends with both stellar mass and bar size, it
makes sense to fit for these trends using logistic regression, which
uses the logistic equation to describe the probability of an object
(such as a barred galaxy) having a binomial characteristic (presence or
absence of an inner bar or nuclear ring) as a function of one or more
parameters. This process uses all the individual data points directly,
and does not depend on any binning scheme (in contrast to the plotted
frequency values in the figures).

More precisely, the probability $P$ of a barred galaxy having a particular
characteristic can be modeled via the logistic equation
\begin{equation}\label{eqn-logistic}
P \; = \; \frac{1}{1 + \mathrm{e}^{-(\alpha \, + \, \sum_{i} \beta_{i} x_{i})}},
\end{equation} 
with $x_{i}$ being the different independent variables (e.g., stellar
mass or bar size). The probability asymptotes to 0 as $x_{i} \rightarrow
-\infty$ and to 1 as $x_{i} \rightarrow +\infty$ (for $\beta_{i} > 0$,
with the reverse behavior for $\beta_{i} < 0$). If a parameter
$x_{i}$ has no relation to the probability, then we would
expect the corresponding slope $\beta_{i}$ to be $\approx 0$. I perform
the fits using the maximum-likelihood approach of the standard \texttt{glm}
function in the R statistical language, which provides an estimate of
\pbeta, the probability for obtaining a slope at least that different
from zero under the null hypothesis where the true slope is zero.

The fitting software also provides Akaike Information Criterion (AIC)
values for each fit, which can be compared for different models fit to
the same data (thus, we can compare AIC values for different models
predicting inner-bar frequencies, or for different models predicting
nuclear-ring frequencies, but not for both sets of models). Lower values
of AIC indicate better fits. One can compute the relative likelihood for
two different models (fit to the same dataset) which differ
by $\Delta$AIC as $\exp(\Delta {\rm AIC}/2)$; in this context, a
difference with significance $P = 0.05$ corresponds to $\Delta{\rm AIC}
\approx -6$, while a 3-$\sigma$ difference is $\Delta{\rm AIC} \approx -11.5$.

Table~\ref{tab:logistic} shows the results of the logistic regression
analysis for both double bars (upper rows) and nuclear rings. In each
case I model the probability as a function of galaxy stellar mass or of
absolute bar size (these are the ``Single-Variable Fits'' in the table).

The very low values of \pbeta{} in all four single-variable fits
strongly suggests that the slopes in each fit are different from zero,
and that there \textit{is} a trend (this is not surprising given how
strong the trends in binned frequencies appear in
Figures~\ref{fig:fDB-NR-vs-logMstar} and
\ref{fig:fDB-NR-vs-logbarsize}). For the case of the double-bar fits, the nearly
identical AIC values (differing by $< 1$) indicate that both bar
size and stellar mass are equally good predictors of whether a galaxy
has a double bar or not. But in the case of nuclear rings, the AIC for
the fit using bar size is $-17.8$ lower than for the fit using stellar
mass; this amounts to a $\sim 3.8\sigma$ significance. So for nuclear
rings, bar size is pretty clearly a better predictor than stellar mass.

The top panel of Figure~\ref{fig:fDB-NR-vs-logMstar} plots the best-fit
logistic curves for double-bar and nuclear-ring presence as a function
of stellar mass. Although the fits are done to all the individual-galaxy
data points and not to the (plotted) binned values, the curves do a
reasonably good job of matching the binned frequencies, except possibly
at the low-mass end, where the curves may overpredict the frequency of
both double bars and nuclear rings.
Figure~\ref{fig:fDB-NR-vs-logbarsize} shows the curves when the
independent variable is bar size; again, the curves do a good job of
matching the binned frequencies.


\begin{table}
\caption{Logistic Regression for DB and NR Presence}
\label{tab:logistic}
\begin{tabular}{@{}lrrrr}
\hline
Variable & $\alpha$ & $\beta$  & \pbeta      & AIC \\
(1)      & (2)      & (3)      & (4)         & (5) \\
\hline

\hline
\multicolumn{5}{c}{Presence of Double Bar: Single-Variable Fits} \\

\logmstarshort   &  $-35.48$   & $3.28$   & $6.2 \times 10^{-6}$   & $133.13$ \\
\logamax   &  $-2.90$   & $3.80$   & $9.2 \times 10^{-6}$   & $133.64$ \\

\multicolumn{5}{c}{Presence of Double Bar: Multiple-variable Fits} \\

\logmstarshort   &  $-25.62$   & $2.24$   & $0.0045$   & $126.94$ \\
\logamax   &   & $2.45$   & $0.0077$   &  \\

\hline
\multicolumn{5}{c}{Presence of Nuclear Ring: Single-Variable Fits} \\

\logmstarshort   &  $-32.39$   & $2.99$   & $1.9 \times 10^{-5}$   & $136.79$ \\
\logamax   &  $-3.60$   & $5.23$   & $3.2 \times 10^{-7}$   & $118.96$ \\

\multicolumn{5}{c}{Presence of Nuclear Ring: Multiple-variable Fits} \\

\logmstarshort   &  $-17.09$   & $1.34$   & $0.094$   & $118.13$ \\
\logamax   &   & $4.31$   & $0.00011$   &  \\

\hline
\end{tabular}

\medskip

Results of logistic regressions: probability of a barred galaxy being
double-barred (upper half of table) or having a nuclear ring (lower
half) as function of (log of) galaxy stellar mass or bar size in kpc. In
the first part of each section (``Single-Variable Fits''), each line
represents a separate logistic regression; in the second part
(``Multiple-Variable Fits''), it is a single fit using both variables. (1)
Galaxy parameter used in fit ($\Mstar{} =$ stellar mass; \amax{} =
semi-major axis of main/outer bar in kpc). (2) Intercept value for fit.
(3) Slope for fit. (4) $P$-value for slope. (5) Akaike Information
Criterion value for fit; lower values indicate better fits.

\end{table}

Since bar size and stellar mass are correlated \citep[e.g.,][]{erwin19},
there is the possibility that the apparent dependence on bar size is a
side effect of a more fundamental dependence on stellar mass -- or vice
versa. (The significantly lower AIC value for the bar-size fit versus
the stellar-mass fit for nuclear-ring presence might be an indication of
the latter.) To test this, I also fit for double-bar or nuclear-ring
presence as a function of \textit{both} stellar mass and bar size
simultaneously; these are the ``Multiple-Variable Fits'' in
Table~\ref{tab:logistic}. For double bars, the fit using both variables
is better than the best single-variable fit ($\Delta$AIC $\approx
-6.2$), with both variables about equally significant. For nuclear
rings, the multi-variable fit is only marginally better than the
single-variable fit using bar size, and the \logmstar{} slope is not
significantly different from zero.

\begin{figure}
\hspace*{-4.5mm}\includegraphics[scale=0.47]{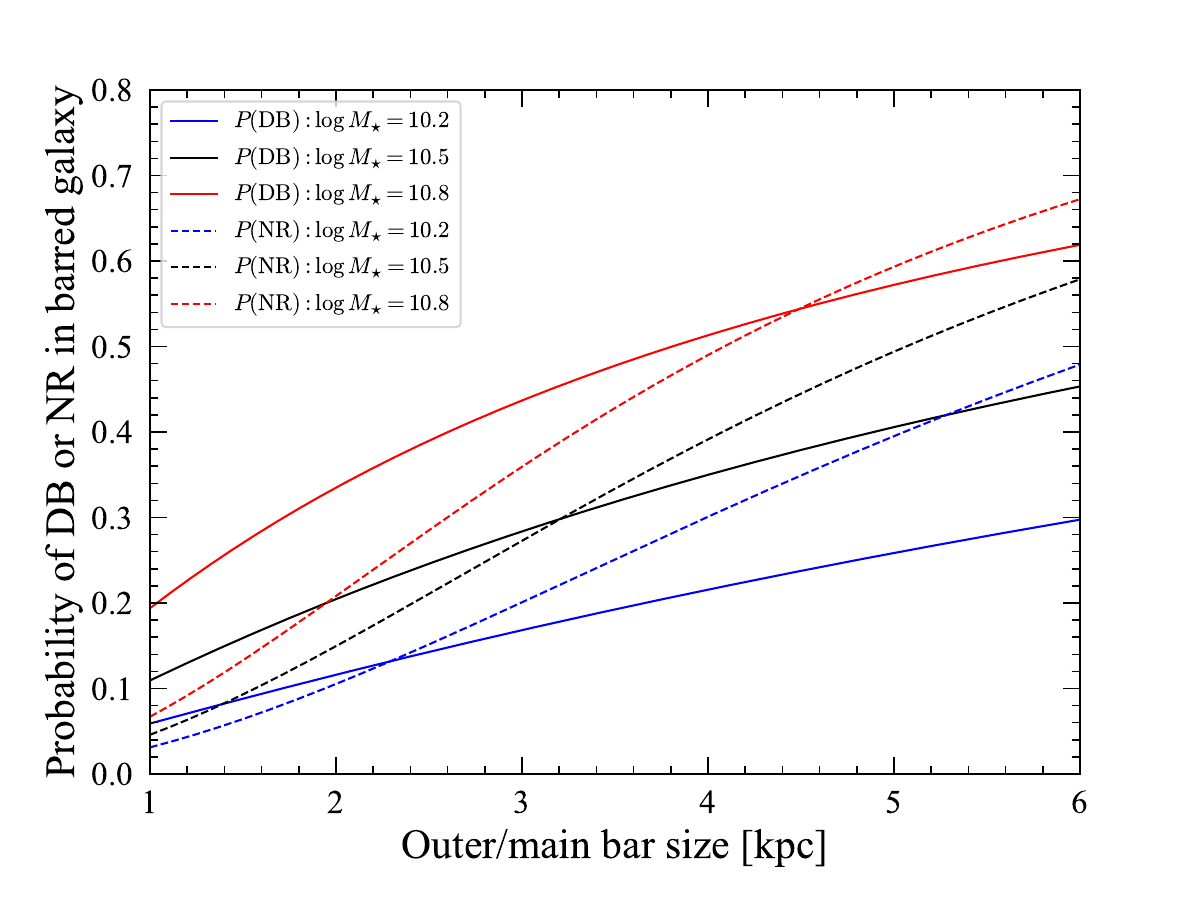}

\caption{Characteristics of the multi-variable logistic fits for
double-bar probability and nuclear-ring probability. Shown are
logistic-fit probabilities for a barred galaxy to have an inner bar
(solid lines) or a nuclear ring (dashed lines) as a function of outer or
single bar size in kpc, for each of three stellar-mass values.
Increasing stellar mass increases the probability of double bars and
nuclear rings, as does increasing the bar size. Note the steeper curves
for the nuclear-ring probability, reflecting the stronger dependence on
bar size.} \label{fig:logistic-probs-db-nr-prob-vs-logbarsize}

\end{figure}

Figure~\ref{fig:logistic-probs-db-nr-prob-vs-logbarsize} illustrates the
combined dependence on stellar mass and bar size for both double bars
and nuclear rings as determined by the multi-variable logistic fits. In
each case, the dependence on bar size when galaxy mass is held fixed is
shown by the curves, with individual curves for fixed values of stellar
mass ($\logmstar = 10.2$, 10.5, and 10.8). The dependence of double-bar
or nuclear-ring probability on stellar mass is clear: the curves are
higher for increasing stellar mass. The dependence on bar size is also
clear: each individual curve shows probability increasing with
increasing bar size. The fact that the nuclear-ring-probability curves
are \textit{steeper} reflects the stronger effect of bar size. For
example, if $\logmstar = 10.5$, then increasing the bar size from 2 to 4
kpc increases the probability of an inner bar from 20\% to 35\%,
while the same doubling of bar size increases the probability of a
nuclear ring from 15\% to 39\%. Roughly speaking, doubling the bar
size increases the chances of finding an inner bar by $\sim 75$\%,
and almost \textit{triples} the chances of finding a nuclear ring.

In summary, whether or not a barred galaxy has an inner bar is a strong
function of stellar mass and a (slightly) weaker function of bar size.
Whether it has a nuclear ring is nominally a function of stellar mass -- but
more significantly a function of bar size; it is plausible that the 
dependence of nuclear-ring presence on stellar mass is mostly a side
effect of the correlation between bar size and stellar mass.

\subsection{Trends with Hubble Type}\label{sec:hubble-type}

\begin{figure}
\hspace*{-1.5mm}
\includegraphics[scale=0.69]{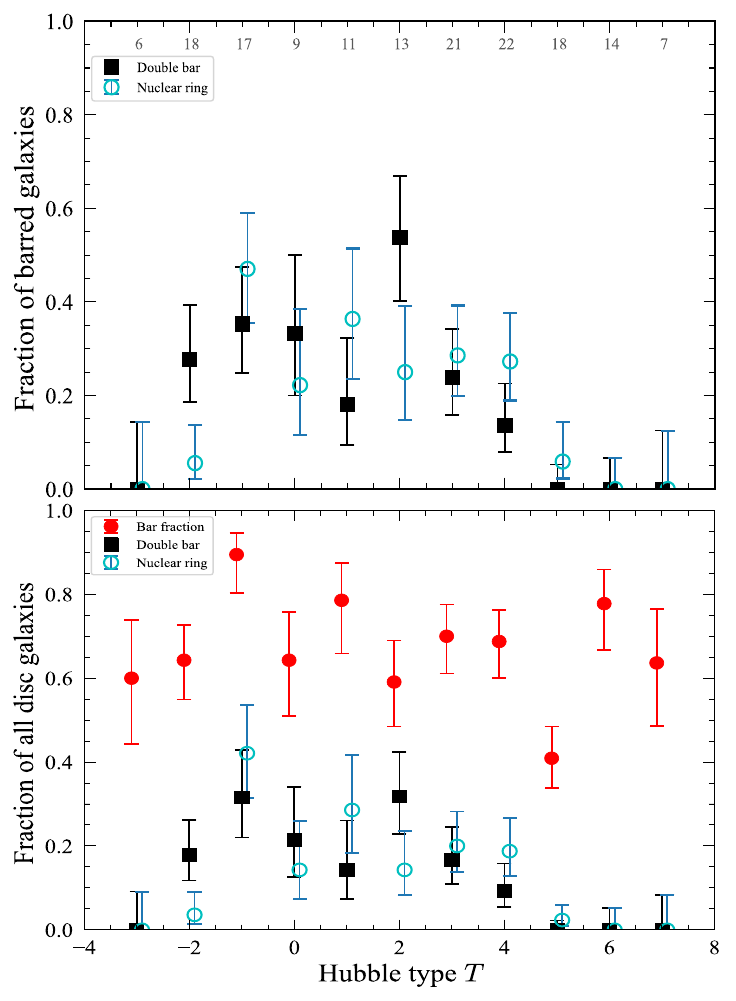}

\caption{As for Figure~\ref{fig:fDB-NR-vs-logMstar}, but now showing frequencies versus 
Hubble type $T$. In contrast to the stellar-mass and bar-size cases 
(Figures~\ref{fig:fDB-NR-vs-logMstar} and \ref{fig:fDB-NR-vs-logbarsize}), 
no clear trends are visible.}
\label{fig:fDB-NR-vs-Htype}
\end{figure}

Figure~\ref{fig:fDB-NR-vs-Htype} show the fraction of inner bars and
nuclear rings as a function of Hubble type. Inner bars are roughly
equally common in (HyperLEDA) Hubble types from $T = -2$ (S$0^{0}$) through 4 (Sbc);
none of the 39 barred Sc--Sd galaxies has an inner bar. Nuclear rings follow
a similar pattern, though they are notably less abundant for $T = -2$
and there is a single nuclear ring in a barred Sc galaxy (NGC~864). It
seems clear that, unlike the case for stellar mass or bar size, inner
bars and nuclear rings do not correlate very strongly with Hubble type,
though their absence in Sc--Sd galaxies might be more than just a stellar-mass
effect.

\section{The Sizes of Inner Bars and Nuclear Rings}\label{sec:sizes} 

Moving on from the question of the mere presence or absence of inner
bars and nuclear rings, there are a number of potentially interesting
characteristics of these structures one might want to
investigate. The most straightforward one is \textit{size}, since this
only requires deprojecting the observed semi-major axis, and can be
directly compared with the size of the outer/sole bar. If we draw on the
general population of (large-scale) bar sizes from \sfourg{}, then we
can investigate questions such as: are there inner bars which are larger
than the bars of single-barred galaxies? Are the outer bars of
double-barred galaxies -- or the host bars of nuclear rings --
systematically larger (or smaller) than the general population of single
bars? How well do the sizes of inner bars or nuclear rings correlate
with host-galaxy characteristics like stellar mass, or outer/sole bar
size?

\begin{figure}
\hspace*{1mm}
\includegraphics[scale=0.7]{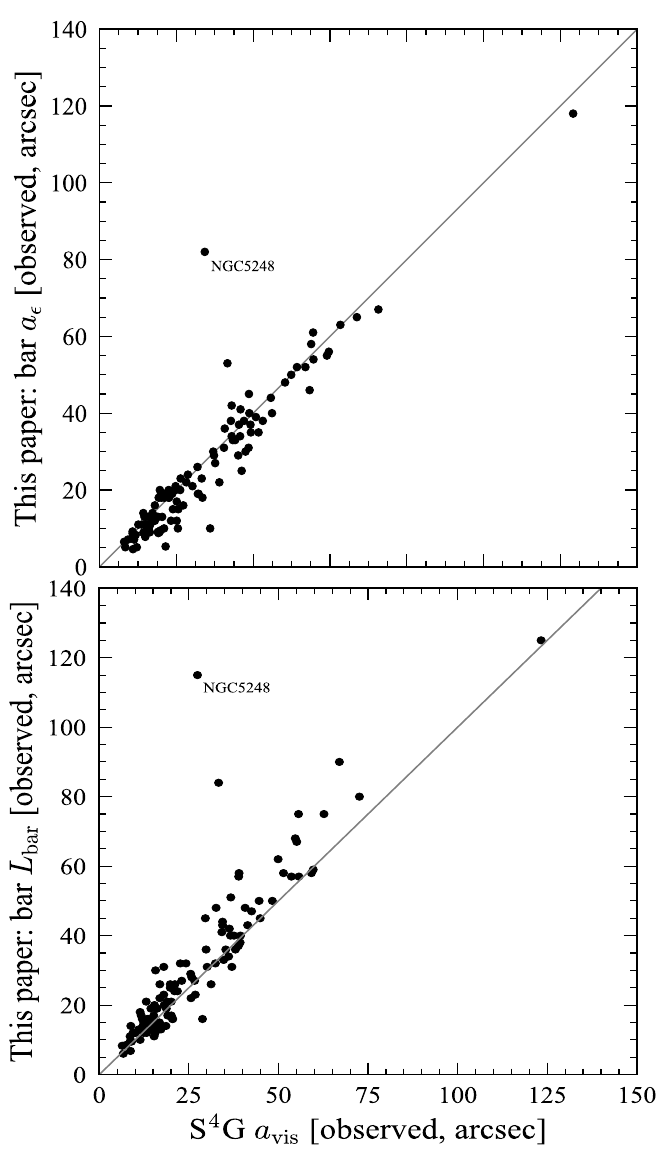}

\caption{Comparison of the two bar-size measurements used in this paper (\amax{}
and \lbar{}) with the \avis{} measurement used for \sfourg{} galaxies by
\citet{herrera-endoqui15}. The outlier is NGC~5248 (see text).} 
\label{fig:barsize-ews-vs-s4g}

\end{figure}

Since I have recorded two size estimates for each bar (\amax{} and
\lbar), it is not immediately obvious which one is the best candidate
for comparison with the much larger set of \sfourg{} bar sizes.
Figure~\ref{fig:barsize-ews-vs-s4g} compares the \sfourg{} \avis{}
measurements from \citet{herrera-endoqui15} with the corresponding
\amax{} and \lbar{} sizes for the 78 bars in common between this study
and Herrera-Endoqui et al.\footnote{For NGC~2681, I use the middle bar
size; see Section~\ref{sec:n2681} for more on this galaxy.} The closest
match to \avis{} is \amax{}, with a median value of $\amax / \avis$ is
0.96; the median of $\lbar / \avis$ is 1.16. The extreme outlier is
NGC~5248, where the ``bar'' as defined by \citet{herrera-endoqui15} is a
slightly boxy oval region inside the large-scale bar identified by
\citet{jogee02a}.

\begin{figure}
\hspace*{-5mm}
\includegraphics[scale=0.47]{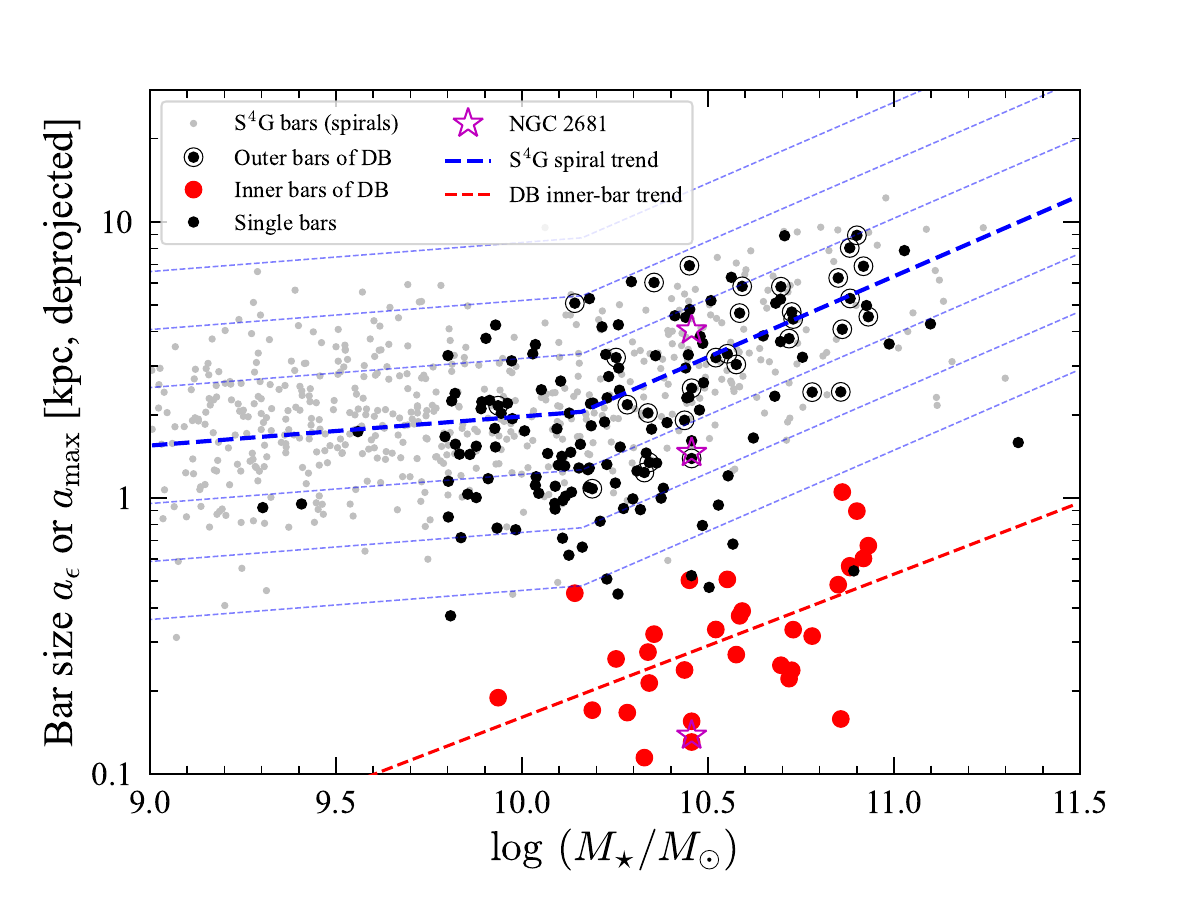}

\caption{Bar size versus galaxy stellar mass. For each galaxy in this
sample, the largest (or only) bar is indicated by a filled black circle;
if the galaxy is double-barred, then the filled circle is surrounded by
a larger open circle. The \textit{inner} bars of double-barred systems
are indicated by red circles. Finally, \avis{} measurements from
\citet{herrera-endoqui15} for the (large-scale) bars of spiral
galaxies in \sfourg{} from \citet{erwin19} that are \textit{not} in this
sample are indicated by small grey circles. The dashed red line indicates
the power-law fit for inner bars. The thick dashed blue line
shows the broken-power-law fit to the bar-size--stellar-mass relation of
\sfourg{} galaxies with $\logmstar = 9$--11 from \citet{erwin19}; the
thinner dashed lines show 1-, 2-, and 3-$\sigma$ boundaries, where
$\sigma = 0.21$ is the standard deviation of \sfourg{} residuals from the fit
in the region $\logmstar = 10$--11. Finally, the sizes of all three
``bars'' in the peculiar galaxy NGC~2681 (Section~\ref{sec:n2681}) are indicated by hollow
magenta stars.}
\label{fig:amax-vs-mstar-with-db}

\end{figure}

Figure~\ref{fig:amax-vs-mstar-with-db} shows the general trend of
large-scale bar sizes from non-sample galaxies in \sfourg{} --
specifically, bars with \avis{} measurements from
\citet{herrera-endoqui15} -- with small gray dots. Also shown is the
bar-size--stellar-mass broken power-law fit from \citet{erwin19}, using
the thick, dashed blue line. The 1-, 2-, and 3-$\sigma$ bounds of this
fit are shown with thinner dashed-blue lines. (These are based on the
standard deviation of the residuals of the fit in the region $\logmstar
= 10$--11; this is $\sigma = 0.21$ in the log of \avis.) The large-scale
bars from this sample are shown (using \amax{} measurements) with larger
filled black circles; these populate the same general trend as the
non-sample \sfourg{} bars, though there is evidence for a population of
unusually small bars in this paper's sample. Those large-scale bars from
this study which are \textit{outer} bars of double-barred galaxies are
encircled in black. Finally, the \textit{inner} bar sizes from this
study are shown with filled red circles.

Several things are apparent from this figure. First, the outer bars of
double-barred systems are entirely typical in size for galaxies with
their stellar masses; they include both relatively large and relatively
small bars. Second, the inner bars are systematically smaller than the
general population of bars. This is, of course, no surprise at all, but
we can quantify this slightly by noting that inner bar sizes are all
below the lower 3-$\sigma$ boundary of the \citet{erwin19} power-law
fit; they are not simply the low-size tail of the general bar-size
distribution. Third, inner-bar size appear to scale with galaxy stellar
mass in a manner similar to (identical to?) the way large-scale bars do.
The dashed red line is a power-law fit to the inner-bar sizes as a
function of stellar mass: the slope is $0.52^{+0.16}_{-0.14}$, consistent
with the high-mass part of the broken power-law fit to \sfourg{} bars
($0.56^{+0.05}_{-0.11}$, from Table~2 of \citealt{erwin19}).

\begin{figure}
\hspace*{-5mm}
\includegraphics[scale=0.47]{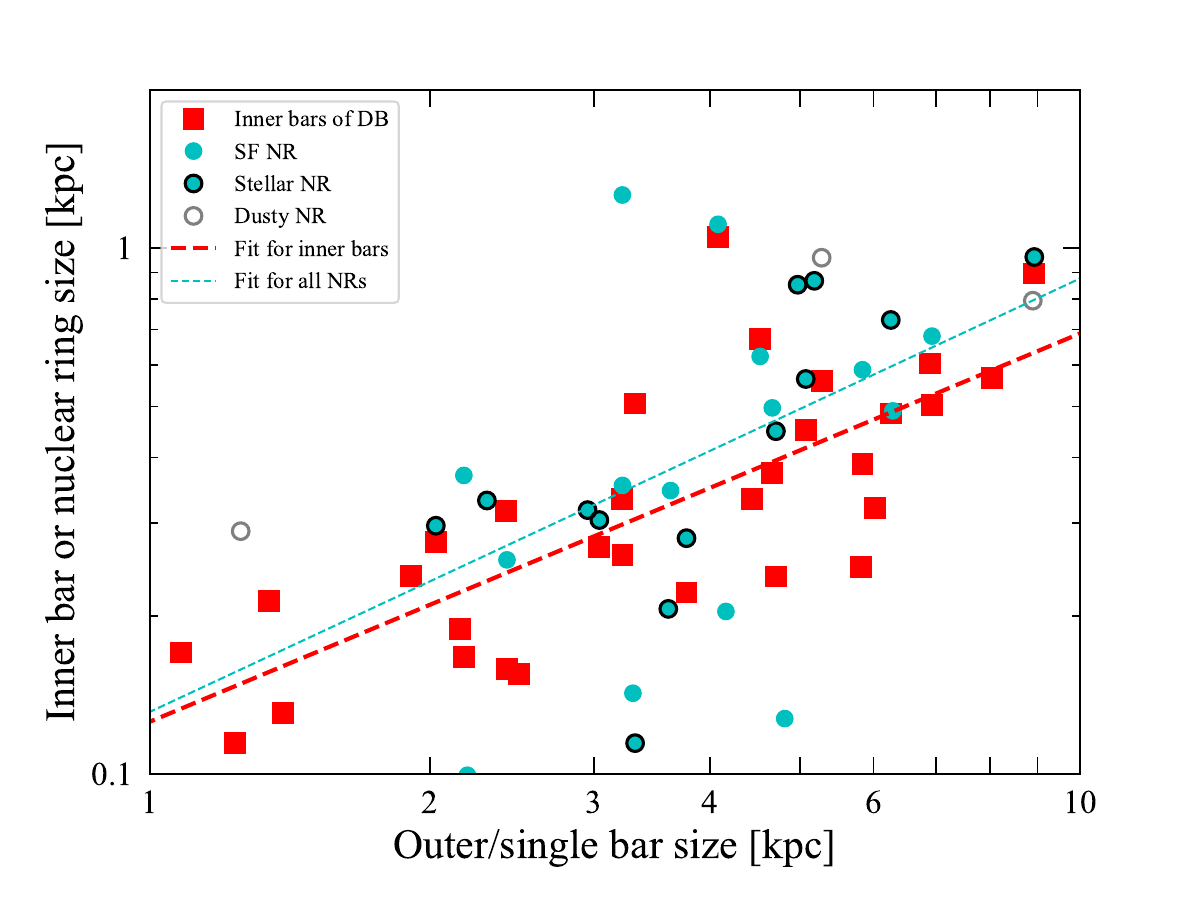}

\caption{Semi-major axes of inner bars (red squares) and nuclear rings
(circles) versus host bar semi-major axis. Dashed lines show power-law fits
for inner bars (thick red line) and nuclear rings (thin cyan line).}
\label{fig:inner-bar-nr-vs-barsize}

\end{figure}

An obvious question at this point is: given that outer-bar sizes are
correlated with galaxy mass (especially for stellar masses $\logmstar
\ga 10.2$), could the size-mass correlation of the \textit{inner} bars
be a side effect of a primary correlation between the sizes of inner and
outer bars? Figure~\ref{fig:inner-bar-nr-vs-barsize} shows that there is
indeed a strong correlation between inner and outer bar sizes, with a
Spearman correlation coefficient $r_{S} = 0.75$ ($P = 1.2 \times
10^{-6}$, compared with $r_{S} = 0.59$ ($P = 0.00047$) for the
correlation with stellar mass. Power-law fits for inner-bar size versus
stellar mass and versus outer-bar size are shown in the upper part of
Table~\ref{tab:size-fits}. The bar-size--bar-size fit has a much lower
AIC ($\Delta{\rm AIC} \approx -241$) and a significantly smaller
\msepred{} (\msepred{} is the mean-squared prediction error, and
measures the difference between the actual sizes of inner bars and their
predicted sizes according to the fit). Thus, it is clear that inner bar
size is primarily correlated with outer bar size, and the correlation
with stellar mass is a side effect.

\begin{figure}
\hspace*{-5mm}
\includegraphics[scale=0.47]{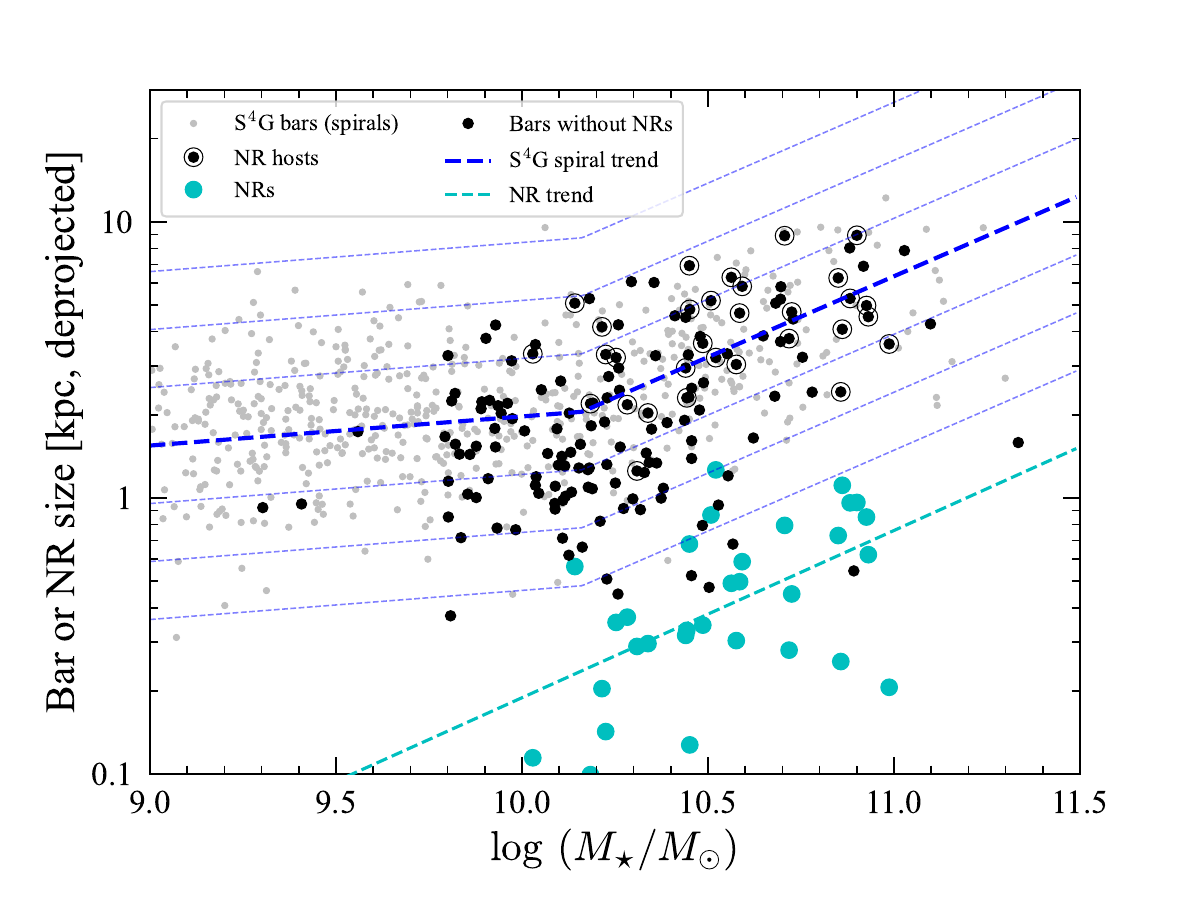}

\caption{As for Figure~\ref{fig:amax-vs-mstar-with-db}, but showing
sizes of nuclear rings (filled cyan circles) and their host bars instead
of double-barred systems; the dashed cyan line shows the power-law
fit to nuclear-ring sizes.} \label{fig:amax-vs-mstar-with-nr}

\end{figure}

\begin{table}
\caption{Fits for IB/NR Size versus Stellar Mass and Host Bar Size}
\label{tab:size-fits}
\begin{tabular}{lccrr}
\hline
Predictor    & $\alpha$  & $\beta$ & AIC & \msepred \\
  (1)        & (2)       & (3)     & (4) & (5)      \\
\hline
\multicolumn{5}{c}{Size of Inner Bar} \\
\logmstarshort   &  $-5.98 \pm 1.50$   & $0.52 \pm 0.14$                 & $630.1$   & $0.045$ \\
\logamax         &  $-0.90 \pm 0.05$   & $0.74 \pm 0.09$                 & $389.2$   & $0.027$ \\
\hline
\multicolumn{5}{c}{Size of Nuclear Ring} \\
\logmstarshort   &  $-6.78^{+2.01}_{-1.97}$   & $0.61 \pm 0.20$          & $979.2$   & $0.070$ \\
\logamax         &  $-0.88^{+0.12}_{-0.16}$   & $0.82^{+0.22}_{-0.16}$   & $1000.1$  & $0.068$ \\
\hline
\end{tabular}

\medskip

Results of linear fits for (logarithm of) inner-bar or nuclear-ring size
as a function of (logarithm of) galaxy stellar mass and host bar size
(outer bar for double-barred galaxies). Parameter uncertainties are
based on 2000 rounds of bootstrap resampling. (1) Predictor variable(s).
(2) Intercept. (3) Slope. (4) Corrected Akaike Information Criterion
value for fit; lower values indicate better fits (comparisons only valid
for same dataset). (5) Mean squared prediction error for log of
inner-bar or nuclear-ring size (kpc), based on 1000 rounds of bootstrap
validation.

\end{table}

What about nuclear rings? In Figure~\ref{fig:amax-vs-mstar-with-nr} we
can see that nuclear rings can be as small as inner bars, but scatter to
larger sizes. Figure~\ref{fig:inner-bar-nr-vs-barsize} shows this is
also true when their sizes are plotted against the size of their host
bars. A numerical comparison confirms that nuclear rings are, on
average, \textit{larger} than inner bars: the median inner-bar size is
315 pc, while the median nuclear-ring size is 370 pc.  This
is also true if we use sizes relative to the host bar (outer bar for
double-barred galaxies): the median inner bar relative size is 0.088
(mean $= 0.10 \pm 0.04$, range = 0.043--0.26), while the median
nuclear-ring size is 0.11 (mean $= 0.12 \pm 0.07$, range = 0.026--0.39).
The results of fitting nuclear-ring size as a function of galaxy mass or
bar sizes, in the lower half of Table~\ref{tab:size-fits}, indicate that
stellar mass is formally a better predictor of nuclear-ring size than
bar size ($\Delta{\rm AIC} = -20.9$), but the \msepred{} values
are basically identical. Although it is tricky to compare fits to
different datasets, the \msepred{} values for the nuclear-ring fits are
much larger than the inner-bar fits (0.070 versus 0.027 in the
case of fits using (outer) bar size as the predictor), and since the
datasets are identical in size, this is probably not a meaningless
comparison. It seems clear that inner bars correlate with the sizes of
their host bars more strongly than nuclear rings do. (Simulations and
theoretical arguments have suggested that nuclear-ring size can depend
on a variety of factors, including the rotation curve, bar pattern
speeds, and the equation of state of the gas; see, e.g.,
\citealt{sormani23} for a discussion and references. Given this, it is
perhaps not surprising that nuclear-ring size does not depend strongly
on just the size of the bar.)

\section{Discussion}

\subsection{Comparison with Previous Studies}

There have been relatively few studies of the demographics of
double-barred galaxies. \cite{erwin02} studied a distance- and
diameter-limited sample of 38 barred S0--Sa galaxies, reporting a
frequency of 25\% for inner bars. An expanded version of this
sample (totalling 76 barred galaxies, including Sab--Sb galaxies)
analyzed in \citet{erwin11} found a frequency of $\sim 30$\%, with
no clear trends with Hubble type. The study of \citet{laine02} used 56
Seyfert galaxies and 56 matched control galaxies, covering Hubble types
S0--Sc. They reported that $17 \pm 4$\% of all the galaxies were double-barred 
(so $28 \pm 5$\% of the barred galaxies had inner bars).


These frequencies are higher than I find for the sample in this paper
(e.g., a $20 \pm 3$\% double-bar fraction for barred galaxies), which can
most likely be explained by a bias towards more massive galaxies --
which are more likely to have inner bars -- in the previous studies, as
well as the exclusion of very late-type spirals, which are both lower
mass and lacking inner bars (Section~\ref{sec:hubble-type}). For
example, the double-bar fraction for barred S0--Sb galaxies in this
paper's sample is $30 \pm 5$\%, the same as reported by \cite{erwin11}
for an earlier (similar but not identical) version of the sample.

Somewhat more attention has been paid to nuclear rings, including the
pioneering study of \citet{buta93} and more recent work by
\citet{knapen05} and \citet{comeron10}. \citet{knapen03} reported a
nuclear-ring fraction of $21 \pm 5$\% for their (diameter-limited)
sample of 57 spirals. This is about twice the frequency in this paper.
As with the double-bar results, this discrepancy can probably be
explained by the fact that a diameter-limited sample will be biased
towards higher-mass galaxies, which are more likely to host nuclear
rings.

\citet{comeron10} reported a ``star-forming'' nuclear-ring fraction of
$20 \pm 2$\% for Hubble types $T = -2$--7. If I exclude the dusty
nuclear rings and use the same range of Hubble types, then the fraction
of SF+stellar nuclear rings is $12 \pm 2$\%. The most plausible reason
for the difference is that the Comer{\'o}n et al.\ sample is magnitude limited
but not distance limited, and so is probably biased in favor of more luminous
and thus more massive galaxies -- where, as we have seen, the fraction of
nuclear rings is higher.

\citet{herrera-endoqui15} looked at nuclear-ring prevalence using
\sfourg{}, which has the advantage of being distance-limited.
Unfortunately, the low resolution of the IRAC1 images, the inability to
identify dusty nuclear rings due to the minimal dust extinction in the
same images, and the absence of unsharp masking as a technique for
finding stellar nuclear rings\footnote{Some stellar nuclear rings may
show up in \citet{buta15} and \citet{herrera-endoqui15} as ``nuclear
lenses'', but this classification is a mixture of nuclear rings, nuclear
discs, and inner bars.} means that most of the nuclear rings found in
this paper are missed by the \citet{buta15} and
\citet{herrera-endoqui15} analysis. Of the 28 \sfourg{} galaxies with
nuclear rings (three more nuclear-ring hosts are S0 galaxies not in
\sfourg), only 9 are listed in \citet{herrera-endoqui15}.

An important takeaway from all of this is that when the frequencies of
particular features (e.g., bars, nuclear rings, etc.) are strongly
dependent on some fundamental galaxy parameter (e.g., stellar mass),
then there is no simple answer to questions like ``what fraction of
galaxies have features X?", and answers can be strongly biased by
selection effects that depend on the same fundamental parameter (galaxy
magnitudes and isophotal diameters are naturally correlated with stellar
mass).

\subsection{Isolated ``Nuclear'' Bars}\label{sec:nuclear-bars} 

Discussions of double-barred galaxies have often involved referring to
the inner bars as ``nuclear bars''. An extension of this is to suggest
that \textit{any} sufficiently small bar is a nuclear bar, even if it is
isolated and without an accompanying outer bar. But how are we to define
what ``sufficiently small'' means? It is not unknown to adopt an
(arbitrary) semi-major axis of 1 kpc
\citep[e.g.,][]{barazza09,melvin14}. But inspection of the population of
(large-scale or outer) bars in local galaxies \citep[e.g.,][]{erwin19}
shows that the general distribution of bar sizes, even for galaxies with
\logmstar{} as high as $\sim 10.1$, smoothly extends to below 1 kpc.
This means it is hard to tell if a strict 1-kpc cutoff is actually a
useful way of identifying qualitatively different (single) bars.

Figure~\ref{fig:amax-vs-mstar-with-db} suggests a possible solution: we
can define a region in the bar-size--\Mstar{} plane which is (mostly)
populated by just the \textit{inner} bars in double-barred galaxies,
using the existing bar-size--\Mstar{} relation of \citet{erwin19}, which
describes the normal range of outer bars and single bars. Thus,
``nuclear bars'' are those with semi-major axes smaller than the 3-$\sigma$
lower limit for the bar-size--stellar-mass relation (the lowest of the
dashed thin blue lines in the figure).

The same figure \textit{also} shows that there are eight \textit{single}
bars which have sizes characteristic of inner bars. Seven of these --
NGC~3982, NGC~3998, NGC~4369, NGC~4041, NGC~4699, NGC~5194 (M51a), and
NGC~5713 -- are in the same mass range as double-barred galaxies
($\logmstar \ga 10.1$); the eighth (UGC~10803) has $\logmstar = 9.81$.
All of these bars are sub-kiloparsec in size, ranging in semi-major axis
from 370--590 pc -- except for NGC~4699 (the most massive galaxy in the
sample), which has a bar with $\amax = 1.26$~kpc. 

Although the classical scenario for double-bar formation involves the
inner bar forming \textit{after} the outer bar, usually from gas inflow
driven by the outer bar (e.g., \citealt{shlosman89};
\citealt{friedli-martinet93}; \citealt{combes94}; see also
\citealt{wozniak15}), a number of $N$-body simulations have inner bars
forming first \citep[e.g.,][]{rautiainen99, debattista07, du15}, as does
the tidal-interaction scenario identified by \citet{semczuk23} from the
IllustrisTNG cosmological simulations. In this context, isolated
nuclear bars could be cases where a (potential) inner bar formed, but then
an outer bar \textit{failed} to form; cases like this were noted
in some of the simulations of \citet{du15}. Alternatively, these could
be cases where the galaxy \textit{was} double-barred, but the
\textit{outer} bar has dissolved in some fashion, leaving just the inner
bar behind. Something like this has also been seen by \citet{du15}, where
some $N$-body simulations evolved from double-barred to single-barred
when the (weak) outer bars dissolved.

\subsection{Pattern Speeds: Relative Position Angles Between Inner and Outer Bars} 

\begin{figure}
\hspace*{-5mm}
\includegraphics[scale=0.47]{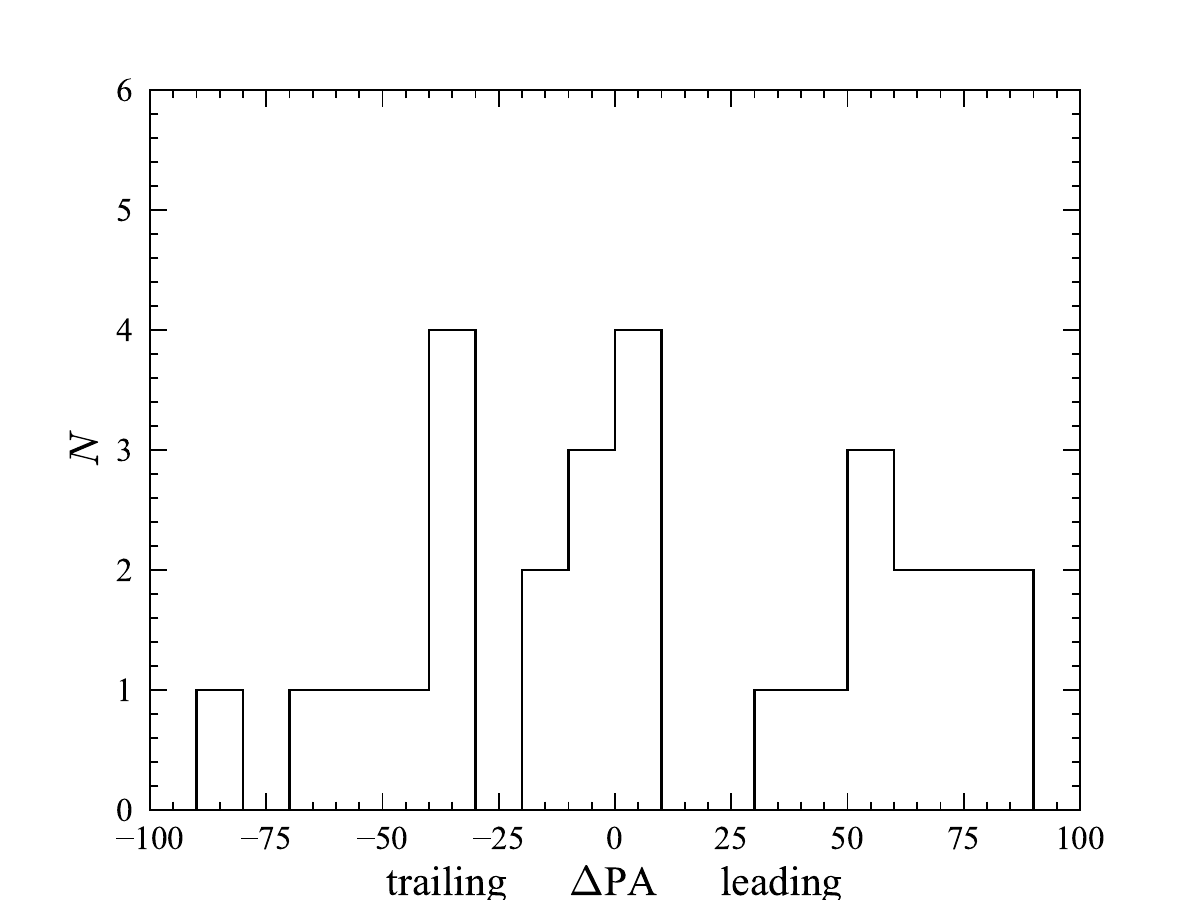}

\caption{Deprojected angles between inner and outer bars of
double-barred galaxies. ''Leading''/''trailing'' refers to whether the
inner bar leads or trails the outer bar, assuming the sense of rotation
suggested by spiral arms and dust lanes in the galaxy.}
\label{fig:db-deltapa-hist}

\end{figure}

A number of studies have looked at the relative orientation of inner and
outer bars in double-barred galaxies. This is because most models of
double bars have assumed -- or demonstrated in the case of $N$-body
models -- that the two bars rotate with different pattern
speeds,\footnote{Note that this does not preclude the pattern
speeds being \textit{coupled} in some fashion, such as resonance overlap
\citep[e.g.,][]{rautiainen99}.} and so should generally have random
relative position angles, as has generally been observed \citep[see,
e.g., the review in][]{erwin11}. In Figure~\ref{fig:db-deltapa-hist} I
show the distribution of relative position angles (PA of inner bar $-$
PA of outer bar, deprojected) for the 31 double bars in this sample.
There does not appear to be a particular favored relative PA or range of
PAs, or a disfavored one; a Kolmogorov-Smirnov test shows no evidence
that the distribution is inconsistent with an underlying uniform one ($P
= 0.39$). Not surprisingly, then, the distribution is consistent with
previous studies and supports the independent-pattern-speed model.

\subsection{The Peculiar Case of NGC~2681: Triple-Barred Galaxy?}\label{sec:n2681}

\citet{wozniak95} noted the curious case of three galaxies they
suggested were, on the basis of ellipse fits to ground-based images,
\textit{triple}-barred, with three nested bars. \citet{erwin99} used
\textit{HST} images to show that the apparent ellipse-fit signatures for two of
those galaxies (NGC~3945 and NGC~4371) were distorted by the combination
of high inclination and large nuclear discs or rings, so that one of the
galaxies was merely double-barred and the other had only a single bar.
But they also argued that a different, face-on galaxy, NGC~2681, was
perhaps genuinely triple-barred. (See also the discussion of this galaxy
in \citealt{erwin-sparke03}.) \cite{laine02} claimed to find two triple-barred
galaxies, but \citet{erwin04} argued that Mrk~573 was only double-barred
and NGC~5033 was probably unbarred.

Both NGC~3945 and NGC~4371 are part of this paper's sample, and remain
stubbornly double- and single-barred, respectively.\footnote{The third
of Wozniak et al.'s triple-barred galaxies was NGC~7187, which is not
part of this paper's sample (and which unfortunately still lacks
\textit{HST} imaging).} NGC~2681 is also in this sample, and remains a
curious, ambiguous case. Figure~\ref{fig:amax-vs-mstar-with-db} shows
the semi-major axes of all three ``bars'' in the context of double- and
single-barred galaxies. Comparison with other double-barred galaxies
strongly suggests the middle and inner bar form a pair that is
reasonably typical of double-barred systems (size ratio = 0.094). This
makes the outermost bar the peculiar component. (The innermost/outermost
size ratio is 0.034, almost half the size of any of the unambiguous
double-barred galaxies.)

\section{Summary}

I have presented the results of a search for inner bars and nuclear rings
in a volume- and mass-limited sample of 155 barred S0--Sd galaxies,
located at distances $< 30$ Mpc with stellar masses $\logmstar > 9.75$.
An equal fraction ($20 \pm 3$\%) of the galaxies have an inner
bar (making them double-barred) and/or a nuclear ring; $11^{+3}_{-2}$\%
of the galaxies have \textit{both}.

The frequency of both inner bars and nuclear rings is a strong function
of stellar mass: more massive galaxies are more likely to have either an
inner bar or a nuclear ring (or both). 47\% of the barred
galaxies with stellar masses $\logmstar > 10.5$ are double-barred, and
45\% have nuclear rings; approximately one-third have both
an inner bar and a nuclear ring.

However, the \textit{strongest} determinant of nuclear-ring presence is
actually the linear size (semi-major axis in kpc) of the host bar.
Approximately 40\% of bars with semi-major axes $> 2$~kpc have nuclear
rings, while barely $\sim 1$\% of smaller bars do. The dependence of
nuclear ring presence on stellar mass appears to be simply a side effect
of the general correlation between (large-scale) bar size and stellar
mass \citep[e.g.,][]{erwin19}. For inner bars, (outer) bar size is a strong
determinant, but does not dominate over stellar mass as it does for
nuclear rings.

The \textit{size} of inner bars correlates with stellar mass, but
correlates more strongly with outer-bar size; inner bars have a median
semi-major axis 0.09 times that of their host (outer) bars, with the
relative sizes ranging from 0.043 to 0.26. Nuclear rings also
scale with host bar size, but the correlation is weaker and the scatter
in size is larger (median ring-radius = 0.11 of bar size, ranging from
0.026 to 0.39).

The correlation of both outer/sole bars and inner bars with stellar mass
enables one to operationally define ``nuclear bars'' as a function of
stellar mass: bars with sizes more than 3-$\sigma$ below the general
bar-size--stellar-mass relation of \citet{erwin19}. This makes it
possible to identify eight galaxies with nuclear bars but no large-scale
bars; these might represent galaxies which formed a potential inner
bar but then failed to form an accompanying outer bar, or which were
originally double-barred but have had their outer bar weaken and
dissolve.

Finally, I note that given that the Milky Way is an Sbc spiral with a
stellar mass of $\logmstar \sim 10.5$ \citep{bland-hawthorn16} and a bar
semi-major axis of $\sim$ 4~kpc \citep{wegg15},\footnote{Wegg et al.\
suggest a maximum bar semi-major axis of 5~kpc; since the fits in this
paper use the semi-major axis of maximum ellipticity \amax{} as the bar
size, I adopt $0.8 \times 5 = 4$~kpc, based on the finding that \amax{}
is typically $\sim 0.8$ times the maximum-size estimate of \lbar{} in
\citet{erwin05b}.} we can use the logistic-regression results of
Section~\ref{sec:logistic} and the fits of Section~\ref{sec:sizes}
to predict a 41\% chance of there being an inner bar (with a most
probable radial size of $\sim 350$ pc), and a 39\% chance of
there being a nuclear ring (with little in the way of size constraints).

\section*{Acknowledgments}

This work has benefitted from comments by and conversations with Lia
Athanassoula, Min Du, Isaac Shlosman, Witold Maciejewski, Kanak Saha,
Fran\c{c}oise Combes, John Beckman, Stuart Robert Anderson, Mattia
Sormani, and especially Victor Debattista. I also thank Preben
Grosb{\o}l for supplying near-IR images of several galaxies, and the
anonymous referee for a number of helpful comments.

This research is based on observations made with the NASA/ESA
\textit{Hubble Space Telescope}, obtained from the data archive at the
Space Telescope Institute.  STScI is operated by the association of
Universities for Research in Astronomy, Inc.\ under the NASA contract
NAS 5-26555.

This work is based in part on observations made with the
\textit{Spitzer} Space Telescope, obtained from the NASA/IPAC Infrared
Science Archive, both of which are operated by the Jet Propulsion
Laboratory, California Institute of Technology under a contract with the
National Aeronautics and Space Administration. This paper also makes use of
data obtained from the Isaac Newton Group Archive which is maintained as
part of the CASU Astronomical Data Centre at the Institute of Astronomy,
Cambridge.

The Legacy Surveys consist of three individual and complementary
projects: the Dark Energy Camera Legacy Survey (DECaLS; Proposal ID
\#2014B-0404; PIs: David Schlegel and Arjun Dey), the Beijing-Arizona Sky
Survey (BASS; NOAO Prop. ID \#2015A-0801; PIs: Zhou Xu and Xiaohui Fan),
and the Mayall z-band Legacy Survey (MzLS; Prop. ID \#2016A-0453; PI:
Arjun Dey). DECaLS, BASS and MzLS together include data obtained,
respectively, at the Blanco telescope, Cerro Tololo Inter-American
Observatory, NSF’s NOIRLab; the Bok telescope, Steward Observatory,
University of Arizona; and the Mayall telescope, Kitt Peak National
Observatory, NOIRLab. The Legacy Surveys project is honoured to be
permitted to conduct astronomical research on Iolkam Du’ag (Kitt Peak),
a mountain with particular significance to the Tohono O’odham Nation.

NOIRLab is operated by the Association of Universities for Research in
Astronomy (AURA) under a cooperative agreement with the National Science
Foundation.

This project used data obtained with the Dark Energy Camera (DECam),
which was constructed by the Dark Energy Survey (DES) collaboration.
Funding for the DES Projects has been provided by the U.S. Department of
Energy, the U.S. National Science Foundation, the Ministry of Science
and Education of Spain, the Science and Technology Facilities Council of
the United Kingdom, the Higher Education Funding Council for England,
the National Center for Supercomputing Applications at the University of
Illinois at Urbana-Champaign, the Kavli Institute of Cosmological
Physics at the University of Chicago, Center for Cosmology and
Astro-Particle Physics at the Ohio State University, the Mitchell
Institute for Fundamental Physics and Astronomy at Texas A\&M
University, Financiadora de Estudos e Projetos, Funda{\c{c}}{\~a}o
Carlos Chagas Filho de Amparo, Financiadora de Estudos e Projetos,
Funda{\c{c}}{\~a}o Carlos Chagas Filho de Amparo {\`a} Pesquisa do
Estado do Rio de Janeiro, Conselho Nacional de Desenvolvimento
Cient{\'i}fico e Tecnol{\'o}gico and the Minist{\'e}rio da Ci{\^e}ncia,
Tecnologia e Inova{\c{c}}o{\~e}s, the Deutsche Forschungsgemeinschaft and the
Collaborating Institutions in the Dark Energy Survey. The Collaborating
Institutions are Argonne National Laboratory, the University of
California at Santa Cruz, the University of Cambridge, Centro de
Investigaciones Energ{\'e}ticas, Medioambientales y
Tecnol{\'o}gicas-Madrid, the University of Chicago, University College
London, the DES-Brazil Consortium, the University of Edinburgh, the
Eidgen{\"o}ssische Technische Hochschule (ETH) Z{\"u}rich, Fermi
National Accelerator Laboratory, the University of Illinois at
Urbana-Champaign, the Institut de Ciencies de l’Espai (IEEC/CSIC), the
Institut de F{\'i}sica d’Altes Energies, Lawrence Berkeley National
Laboratory, the Ludwig Maximilians Universit{\"a}t M{\"u}nchen and the
associated Excellence Cluster Universe, the University of Michigan,
NSF’s NOIRLab, the University of Nottingham, the Ohio State University,
the University of Pennsylvania, the University of Portsmouth, SLAC
National Accelerator Laboratory, Stanford University, the University of
Sussex, and Texas A\&M University.

BASS is a key project of the Telescope Access Program (TAP), which has
been funded by the National Astronomical Observatories of China, the
Chinese Academy of Sciences (the Strategic Priority Research Program
``The Emergence of Cosmological Structures'' Grant \# XDB09000000), and the
Special Fund for Astronomy from the Ministry of Finance. The BASS is
also supported by the External Cooperation Program of Chinese Academy of
Sciences (Grant \# 114A11KYSB20160057), and Chinese National Natural
Science Foundation (Grant \# 11433005).

The Legacy Survey team makes use of data products from the Near-Earth
Object Wide-field Infrared Survey Explorer (NEOWISE), which is a project
of the Jet Propulsion Laboratory/California Institute of Technology.
NEOWISE is funded by the National Aeronautics and Space Administration.

The Legacy Surveys imaging of the DESI footprint is supported by the
Director, Office of Science, Office of High Energy Physics of the U.S.
Department of Energy under Contract No. DE-AC02-05CH1123, by the
National Energy Research Scientific Computing Center, a DOE Office of
Science User Facility under the same contract; and by the U.S. National
Science Foundation, Division of Astronomical Sciences under Contract No.
AST-0950945 to NOAO.

The Siena Galaxy Atlas was made possible by funding support from the
U.S. Department of Energy, Office of Science, Office of High Energy
Physics under Award Number DE-SC0020086 and from the National Science
Foundation under grant AST-1616414.

This research also made use of both the NASA/IPAC Extragalactic Database
(NED) which is operated by the Jet Propulsion Laboratory, California
Institute of Technology, under contract with the National Aeronautics
and Space Administration, and the Lyon-Meudon Extragalactic Database
(HyperLEDA; http://leda.univ-lyon1.fr).

Finally, this research made use of Astropy, a community-developed core
Python package for Astronomy \citep{astropy22}.

\section*{Data availability}

The data underlying this article, along with code for reproducing fits
and figures, are available at \url{https://github.com/perwin/db-nr_paper}
and also at \url{https://doi.org/10.5281/zenodo.10252783}.

\bibliographystyle{mnras}

\appendix

\section{Resolution Effects}\label{app:resolution}

Bars that are too small relative to the resolution of an image can be
hard to detect, and may therefore be missed. If bar size has some
variance \textit{and} is correlated with a galaxy parameter of interest,
such as stellar mass, then a resolution limit can interact with this to
produce (or exaggerate) an apparent decrease of bar frequency with
change in galaxy parameter: intrinsically smaller bars become
probabilistically harder to detect, and so galaxies that tend to have
\textit{smaller} bars will appear to have \textit{fewer} bars. Indeed,
just such an effect was suggested as the cause for the difference in the
observed dependence of bar fraction on galaxy mass between \sfourg{} and
more distant SDSS samples in \citet{erwin18}. Since there is evidence
that inner bars have, like large-scale bars, a dependence on galaxy mass
(Figure~\ref{fig:amax-vs-mstar-with-db}), the possibility of resolution
effects is something that should be checked.

A plausible rule of thumb is that a bar should have a semi-major axis at
least twice the FWHM of the point-spread function (PSF) of an image
\citep[e.g.,][]{aguerri09,erwin18,liang23} in order for it to be reliably detected.
For a given FWHM in angular size, galaxies at larger distances will have
an effective spatial resolution (in pc) worse than galaxies at smaller
distances. In addition, this survey relies on a heterogenous set of
observations with a wide range different angular resolutions, ranging
from optical \textit{HST} images (FWHM $\approx 0.07\arcsec$) to
\textit{Spitzer} IRAC1 images (FWHM $\approx 1.6\arcsec$). Given all
this, it is worth examining how much varying spatial resolutions could
affect the detectability of inner bars and nuclear rings.

\begin{figure}
\hspace*{-4mm}\includegraphics[scale=0.465]{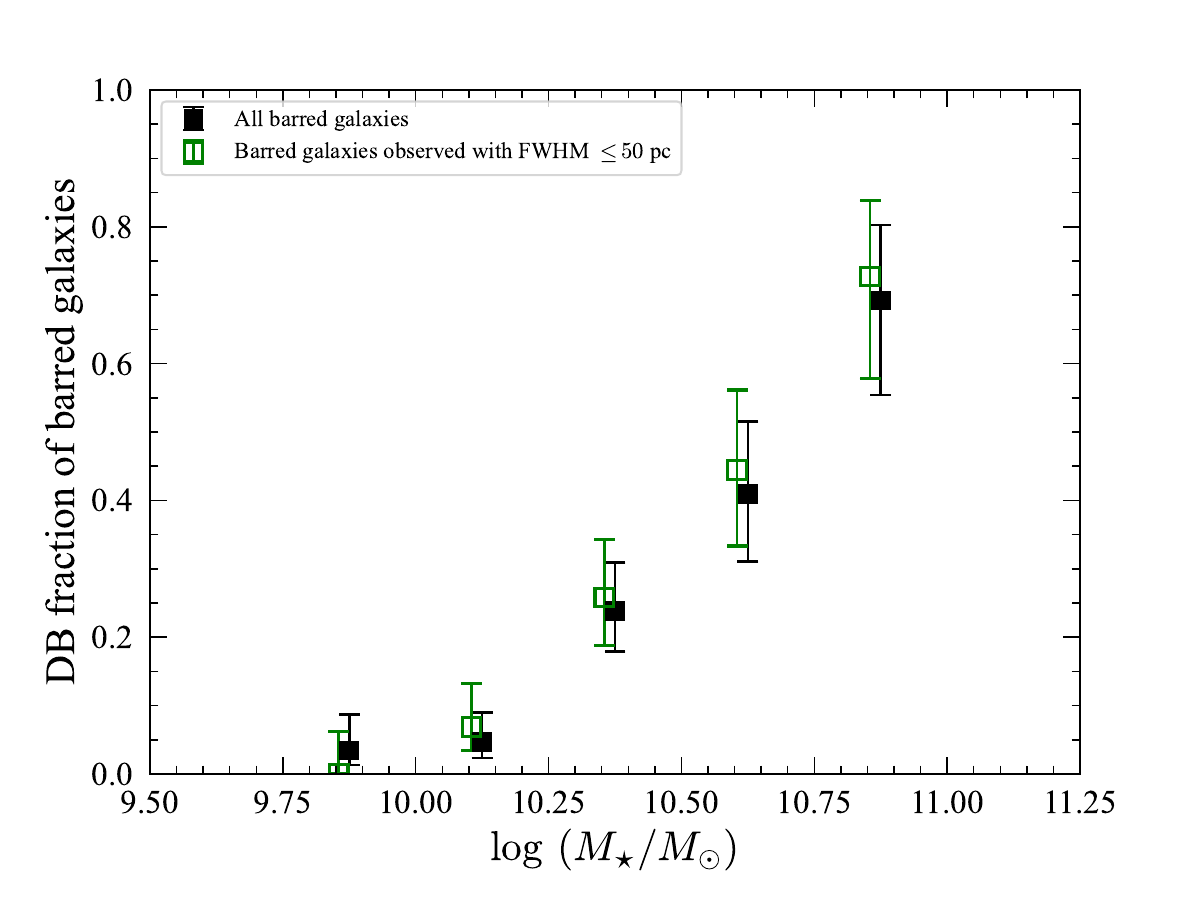}

\caption{As for Figure~\ref{fig:fDB-NR-vs-logMstar}, but showing just
the double-bar fraction for all barred galaxies (black squares) and for
only those barred galaxies with spatial-resolution FWHM better than 50
pc (open green squares) for the images used for inner-bar
identification. The trends are almost identical, suggesting that the
results are not strongly affected by varying image resolution.} \label{fig:fdb-vs-fwhm}

\end{figure}

The smallest inner bars mentioned in the review of \citet{erwin11} have
semi-major axes $\sim 100$ pc (and the smallest bar in \textit{this}
paper has $\amax = 115$ pc), so we should probably consider that our
limit. (Nuclear rings tend to be \textit{larger} than inner bars (e.g.,
Section~\ref{sec:sizes}.) We can therefore consider a galaxy properly
resolved if the spatial FWHM is $< 50$ pc. Figure~\ref{fig:fdb-vs-fwhm}
shows the trend of double-bar fraction versus stellar mass for the
entire sample (filled black squares, as for
Fig.~\ref{fig:fDB-NR-vs-logMstar}), and then for the ``properly
resolved'' subset containing just those galaxies with spatial FWHM $<
50$ pc (open green squares). The double-bar fraction is
\textit{slightly} higher at all masses $\logmstar > 10.25$, which
suggests the survey as a whole \textit{might} be underestimating the
double-bar fraction. But the differences are marginal, and it is
plausible to say that two trends are identical within the uncertainties.
This provides good evidence that the overall analysis in this paper is
\textit{not} strongly affected by resolution issues.

\section{Discarded Galaxies}\label{app:discarded}

A total of 25 galaxies were discarded from the initial sample due to
being too highly inclined or even edge-on (NGC~3912, NGC~4435, NGC~4606,
NGC~4638, NGC~5103, NGC~5493, NGC~5574, NGC~5667, UGC~9215, and PGC
27316); distorted by ongoing interactions or post-merger states, which
can make it difficult to properly deproject bar or ring sizes (NGC~274,
NGC~275, NGC~2655, NGC~2685, NGC~3227, NGC~3310, NGC~3395, NGC~3414,
NGC~4382, NGC~4567, NGC~5195, NGC~5953, NGC~7465, and UGC~5814), or too
close to bright stars (IC~630).

Three galaxies were removed for having measured distances outside the 30
Mpc limit: NGC~3021 \citep[Cephied distance = 31.6 Mpc][]{riess16};
NGC~3504 \citep[surface brightness fluctuation distance = 34.4
Mpc;][]{jensen21}; and NGC~4386 \citep[SBF distance = 30.6
Mpc][]{jensen21}. Six more had group distances (i.e., membership in
groups with at least three direct distance measurements) in
\citet{kourkchi-tully17} outside 30 Mpc: IC~3102 (31.4 Mpc), NGC~524
(31.2 Mpc), NGC~4269 (31.4 Mpc), NGC~4292 (31.4 Mpc), NGC~4378 (31.4
Mpc), and NGC~5371 (32.0 Mpc). An additional galaxy (NGC~4389) had
a group distance small enough (7.3 Mpc) so that its revised stellar mass
($\logmstar = 9.21$) put it below the sample cutoff. Finally, NGC~3730
has no distance in \sfourg, and the NED and HyperLEDA redshifts strongly
disagree, so I removed it as well.

\section{Notes on Individual Barred Galaxies}\label{app:barred} 

Notes and measurements for the following galaxies can be found in
\citet{erwin-sparke03}: IC~676, NGC~718, NGC~936, NGC~1022, NGC~2681
\citep[see also][]{erwin99}, NGC~2859 \citep[see
also][]{erwin15-composite}, NGC~2950, NGC~2962, NGC~3185, NGC~3412,
NGC~3729, NGC~3941, NGC~3945 \citep[see
also][]{erwin99,erwin03-id,erwin15-composite}, NGC~4143, NGC~4203,
NGC~4245 (see also below), NGC~4314, NGC~4386, NGC~4624 (as
``NGC~4665''), NGC~4691, NGC~5701, and NGC~7280. \citet{gutierrez11}
provides details for NGC~3599 and NGC~3998.

For most other galaxies, the ``disc orientation'' (PA and inclination) is
generally based on the shape of disc isophotes (i.e., from ellipse fits
to images) well outside the bar, assuming circularity and an intrinsic
disc flattening $c/a = 0.2$. The specific image or images used for
each galaxy is given in Table~\ref{tab:sample}. 
Other sources (used when, e.g., outer-disc isophotes were too low in
S/N, did not have consistent, unvarying orientations due to strong
spiral arms, etc.) are listed below for specific galaxies.

\textbf{NGC 514}: Disc orientation from \textit{HST} WFC3-IR F160W
image (Proposal ID 16407, PI Grauer).

\textbf{NGC 864}: Disc orientation from \ha{} velocity field \citep{epinat08a}.
This galaxy has the smallest nuclear ring (deprojected $a = 99$~pc) in the sample.

\textbf{NGC 1068 (M77)}: See \citet{erwin04} for discussion of evidence
for both inner and outer bars, and \citet{erwin15-composite} for
detailed morphological and kinematic analysis. This galaxy has the
largest inner bar in the sample (deprojected $\amax = 1.04$~kpc).

\textbf{NGC 1637}: Disc orientation from gas velocity fields in \citet{williams21}.

\textbf{NGC 2608}: Disc orientation from \textit{Spitzer} IRAC1 image and 2D
decomposition of \citet{salo15}.

\textbf{NGC 2681}: See Section~\ref{sec:n2681}.

\textbf{NGC 2787}: See \citet{erwin-sparke03}. The morphology
and stellar kinematics inside the bar were analyzed in detail in \citet{erwin03-id};
however, the ``inner disc'' they identified is more likely to be the projected
B/P structure of the bar. The two ``dust'' rings mentioned in \citet{comeron10}
are associated with off-plane gas \citet{erwin03-id,silchenko04}.

\textbf{NGC 2968}: I use the HyperLEDA Hubble-type (SBa) rather than the
RC3 I0 classification.

\textbf{NGC 3031 (M81)}: Although this galaxy was classified as unbarred
in \citet{rc3}, it is actually double-barred \citep{gutierrez11}. The
inner bar was first identified (as a ``minibar'') by
\citet{elmegreen95}; the weak outer bar was described by
\citet{gutierrez11} and \citet{erwin-debattista13}.

\textbf{NGC 3368 (M96)}: See \citet{nowak10} and \citet{erwin15-composite}.

\textbf{NGC 3458}: This galaxy has the smallest inner
bar in the sample (deprojected $\amax = 115$~pc).

\textbf{NGC 3489}: See \citet{erwin-sparke03}, \citet{nowak10}, and \citet{erwin15-composite}.

\textbf{NGC 3626}: Although classified as unbarred by \citet{rc3}, \citet{laurikainen05}
and \citet{gutierrez11} found it to be double-barred.

\textbf{NGC 3726}: Disc orientation from \hi{} velocity field \citep{van-eymeren11}.

\textbf{NGC 3982}: The ``nuclear ring'' noted by \citet{comeron10} lies
outside the sole bar in this galaxy, and so I consider it an inner ring 
instead.

\textbf{NGC 3992 (M109)}: Disc orientation from \hi{} velocity field \citep{verheijen01}.

\textbf{NGC 4041}: Small-scale (``nuclear'') single bar in nominally unbarred
galaxy \citep[e.g.,][]{comeron14}.

\textbf{NGC 4051}: Disc orientation from \hi{} velocity field \citep{liszt95}.

\textbf{NGC 4102:} The ``nuclear ring'' noted by \citet{comeron10} is,
in \textit{HST} near-IR images, a complex, possibly off-center structure
that is not clearly ringlike, so I do not classify it as a nuclear ring.

\textbf{NGC 4151}: Disc orientation from \hi{} velocity field \citep{pedlar92}.

\textbf{NGC 4221}: This is the lowest-mass nuclear-ring host galaxy in
the sample.

\textbf{NGC 4245}: The \textit{HST} WFC3-IR image shows inner bar not
visible in optical, ground-based images analyzed by
\citet{erwin-sparke03}.

\textbf{NGC 4250}: Disc orientation from \citet{salo15}.

\textbf{NGC 4303}: Disc orientation from \citet{schinnerer02}.

\textbf{NGC 4319}: Disc orientation from analysis of archival $R$-band images
from the Isaac Newton Group.

\textbf{NGC 4321 (M100)}: Disc orientation from \hi{} velocity field \citep{haan08}.

\textbf{NGC 4371}: Disc orientation from \citet{erwin08}. 

\textbf{NGC 4608}: See \citet{erwin21}.

\textbf{NGC 4643}: See \citet{erwin21}, where the stellar nuclear ring
suggested by \citet{erwin04} is demonstrated to be a nuclear disc with a
broken-exponential profile.

\textbf{NGC 4699}: See \citet{erwin15-composite}.

\textbf{NGC 4713}: Disc orientation from \ha{} velocity field \citep{epinat08a}.

\textbf{NGC 4725}: Disc orientation from a combination of ellipse fits to IRAC1 image
and \hi{} velocity field \citep{ponomareva16}.

\textbf{NGC 4736 (M94)}: Disc orientation from stellar and gas
kinematics in \citet{moellenhoff95} and \citet{van-driel96}. This galaxy
has the largest nuclear ring (deprojected $a = 1.26$~kpc) in the sample.

\textbf{NGC 4750}: See \citet{gutierrez11}. This was suggested as
\textit{possible} double-bar by \citet{erwin04}, with the outer bar being
uncertain, primarily due to the low S/N of the 2MASS near-IR images. The
\textit{Spitzer} IRAC1 image shows very clearly that the spiral arms lie
in an oval region misaligned with respect to the outer disc. This makes
the outer bar somewhat similar to the bar in NGC~5248.

\textbf{NGC 4941:} This was suggested as possible but unconfirmed double-bar in
\citet{erwin04}, due to uncertainty about the existence of the outer
bar. Evidence for the latter's reality includes offset gas lanes in
detected in CO by \citet{stuber23} and also in ionized-gas emission (T.
Kolcu, private comm.).

\textbf{NGC 5194 (M51a)}: Disc orientation from \hi{} velocity field
\citep{tamburro08}. \citet{comeron10} note the existence of a
star-forming``nuclear'' ring with $a = 16.1\arcsec$; since this is
\textit{larger} than the only bar in the galaxy, I do not consider it a
true nuclear ring.

\textbf{NGC 5248}: I consider the (very) large-scale structure identified by \citet{jogee02a}
to be the bar, rather than the (much smaller) ``bar'' reported by \citet{herrera-endoqui15}.
For simplicity, I only list the larger of the two star-forming nuclear rings (the smaller,
with $a \approx 1.5\arcsec$, would be the smallest nuclear ring in terms of size relative
to the bar).

\textbf{NGC 5377}: \citet{erwin-sparke03} listed \textit{two} nuclear rings in this
galaxy. The first, a larger dusty ring, is unambiguous and is the one used here.
The second (a ``blue'' nuclear ring) is perhaps more of a blue nuclear disc, marked
by weak inner dust lanes.

\textbf{NGC 5457 (M101)}: Disc orientation from inner part of H I velocity field
\citep{bosma81a}.

\textbf{NGC 5480}: Disc orientation from CO velocity field \citep{levy18}.

\textbf{NGC 5770}: This is the lowest-mass double-barred galaxy in the sample.

\textbf{NGC 5806}: Disc orientation from INT-WFC $r$ images \citep{erwin08}.

\textbf{NGC 5850}: Disc orientation from \hi{} velocity field \citep{higdon98a}.

\textbf{NGC 5964}: Disc orientation from \hi{} \citep{hewitt83} and \ha{} \citep{hernandez05a}
velocity fields.

\textbf{NGC 6412}: Disc orientation from \ha{} velocity field \citep{epinat08a}.

\textbf{NGC 7741}: Disc orientation from IRAC1 and \ha{} velocity field \citep{fathi09}.

\textbf{NGC 7743}: See \citet{erwin-sparke03}. An updated discussion of
the orientation of this galaxy can be found in \citet{davies14}.

\textbf{PGC 12633}: Disc orientation from IRAC1 and SGA images.

\bsp	
\label{lastpage}
\end{document}